\documentclass[sn-mathphys,Numbered,iicol]{sn-jnl}% Math and Physical Sciences Reference Style
\usepackage{graphicx}%
\usepackage{multirow}%
\usepackage{amsmath,amssymb,amsfonts}%
\usepackage{amsthm}%
\usepackage{mathrsfs}%
\usepackage[title]{appendix}%
\usepackage{xcolor}%
\usepackage{textcomp}%
\usepackage{manyfoot}%
\usepackage{booktabs}%
\usepackage{etoolbox}
\usepackage{algorithm}%
\usepackage{algorithmicx}%
\usepackage{algpseudocode}%
\usepackage{listings}%
\usepackage{fancyhdr}
\usepackage{csquotes}
\usepackage{bm}

\usepackage[utf8]{inputenc}  % UTF-8 input encoding
\usepackage[T1]{fontenc}     % Type1 fonts
\usepackage{lmodern}         % Improved Computer Modern font
\usepackage{microtype}       % Better handling of typo
\usepackage[scaled]{helvet}  % Scale Helvetica
\usepackage[english]{babel}  % Hyphenation
\usepackage{graphicx}        % Import graphics
\usepackage{xspace}
\usepackage{url} 
\theoremstyle{thmstyleone}%
\theoremstyle{thmstyletwo}%

\theoremstyle{thmstylethree}%

\begin{document}

%%%%%%%%%%%%%%%%%%%%%%%%%%%%%%%%%%%%%%%%%%%%%%%%%%
% These are some new commands that may be useful 
% for paper writing in general. If other newcommands
% are needed for your specific paper, please feel 
% free to add here. 
%
% The currently available commands are organized in: 
% 1) Systems
% 2) Quantities
% 3) Energies and units
% 4) Detectors
% 5) particle species 
%%%%%%%%%%%%%%%%%%%%%%%%%%%%%%%%%%%%%%%%%%%%%%%%%%

% 1) SYSTEMS 
\newcommand{\pp}           {\ensuremath{\mathrm{pp}}\xspace}
\newcommand{\ppbar}        {\mbox{\ensuremath{\mathrm {p\overline{p}}}}\xspace}
\newcommand{\XeXe}         {\ensuremath{\mathrm{\mbox{XeXe}}}\xspace}
\newcommand{\PbPb}         {\ensuremath{\mathrm{\mbox{PbPb}}}\xspace}
\newcommand{\pA}           {\ensuremath{\mathrm{\mbox{pA}}}\xspace}
\newcommand{\pPb}          {\ensuremath{\mathrm{\mbox{pPb}}}\xspace}
\newcommand{\AuAu}         {\ensuremath{\mathrm{\mbox{AuAu}}}\xspace}
\newcommand{\OO}         {\ensuremath{\mathrm{\mbox{OO}}}\xspace}
\renewcommand{\AA}           {\ensuremath{\mathrm{\mbox{AA}}}\xspace}
\newcommand{\dAu}          {\ensuremath{\mathrm{\mbox{dAu}}}\xspace}
\newcommand{\ee}           {\ensuremath{\mathrm{e^{+}e^{-}}}\xspace}
\newcommand{\ep}           {\ensuremath{\mathrm{e^{\pm}p\xspace}}}
\newcommand{\qqbar}        {\ensuremath{\mathrm{q}\overline{\mathrm{q}}}\xspace}

% 2) QUANTITIES 
\newcommand{\s}            {\ensuremath{\sqrt{s}}\xspace}
\newcommand{\snn}          {\ensuremath{\sqrt{s_{\mathrm{NN}}}}\xspace}
\newcommand{\pt}           {\ensuremath{p_{\rm T}}\xspace}
\newcommand{\meanpt}       {$\langle p_{\mathrm{T}}\rangle$\xspace}
\newcommand{\ycms}         {\ensuremath{y_{\rm CMS}}\xspace}
\newcommand{\ylab}         {\ensuremath{y_{\rm lab}}\xspace}
\newcommand{\etarange}[1]  {\mbox{$\left | \eta \right |~<~#1$}}
\newcommand{\yrange}[1]    {\mbox{$\left | y \right |~<~#1$}}
\newcommand{\dndy}         {\ensuremath{\mathrm{d}N_\mathrm{ch}/\mathrm{d}y}\xspace}
\newcommand{\dndeta}       {\ensuremath{\mathrm{d}N_\mathrm{ch}/\mathrm{d}\eta}\xspace}
\newcommand{\avdndeta}     {\ensuremath{\langle\dndeta\rangle}\xspace}
\newcommand{\dNdy}         {\ensuremath{\mathrm{d}N_\mathrm{ch}/\mathrm{d}y}\xspace}
\newcommand{\Npart}        {\ensuremath{N_\mathrm{part}}\xspace}
\newcommand{\Ncoll}        {\ensuremath{N_\mathrm{coll}}\xspace}
\newcommand{\dEdx}         {\ensuremath{\textrm{d}E/\textrm{d}x}\xspace}
\newcommand{\RpPb}         {\ensuremath{R_{\rm pPb}}\xspace}
\newcommand{\Tfo}         {\ensuremath{T_\mathrm{fo}}\xspace}
\newcommand{\Th}         {\ensuremath{T_\mathrm{h}}\xspace}
\newcommand{\de}          {\ensuremath{\mathrm{d}}\xspace}
\newcommand{\Raa}          {\ensuremath{R_\mathrm{AA}}\xspace}
\newcommand{\FFc}{\ensuremath{f(\mathrm{c\rightarrow H_{c}})}\xspace}
\newcommand{\zjet}{\ensuremath{z_{\mathrm{||}}^{\mathrm{ch}}}\xspace}
% 3) ENERGIES, UNITS
\newcommand{\nineH}        {$\sqrt{s}~=~0.9$~Te\kern-.1emV\xspace}
\newcommand{\seven}        {$\sqrt{s}~=~7$~Te\kern-.1emV\xspace}
\newcommand{\twoH}         {$\sqrt{s}~=~0.2$~Te\kern-.1emV\xspace}
\newcommand{\twosevensix}  {$\sqrt{s}~=~2.76$~Te\kern-.1emV\xspace}
\newcommand{\five}         {$\sqrt{s}~=~5.02$~Te\kern-.1emV\xspace}
\newcommand{\twosevensixnn}{$\sqrt{s_{\mathrm{NN}}}~=~2.76$~Te\kern-.1emV\xspace}
\newcommand{\fivenn}       {$\sqrt{s_{\mathrm{NN}}}~=~5.02$~Te\kern-.1emV\xspace}
\newcommand{\LT}           {L{\'e}vy-Tsallis\xspace}
\newcommand{\GeVc}         {\ensuremath{\mathrm{GeV}/c}\xspace}
\newcommand{\MeVc}         {\ensuremath{\mathrm{MeV}/c}\xspace}
\newcommand{\TeV}          {\ensuremath{\mathrm{TeV}}\xspace}
\newcommand{\GeV}          {\ensuremath{\mathrm{GeV}}\xspace}
\newcommand{\MeV}          {\ensuremath{\mathrm{MeV}}\xspace}
\newcommand{\GeVmass}      {\ensuremath{\mathrm{GeV}/c^2}\xspace}
\newcommand{\MeVmass}      {\ensuremath{\mathrm{MeV}/c^2}\xspace}
\newcommand{\lumi}         {\ensuremath{\mathcal{L}}\xspace}
\newcommand{\mub}         {\ensuremath{\mu\mathrm{b}}\xspace}
\newcommand{\Taa}          {\ensuremath{\langle T_\mathrm{AA}\rangle}\xspace}
\newcommand{\Diffs}        {\ensuremath{D_{s}}\xspace}
\newcommand{\vtwo}          {\ensuremath{v_\mathrm{2}}\xspace}
% 4) DETECTORS 
\newcommand{\ITS}          {\rm{ITS}\xspace}
\newcommand{\TOF}          {\rm{TOF}\xspace}
\newcommand{\ZDC}          {\rm{ZDC}\xspace}
\newcommand{\ZDCs}         {\rm{ZDCs}\xspace}
\newcommand{\ZNA}          {\rm{ZNA}\xspace}
\newcommand{\ZNC}          {\rm{ZNC}\xspace}
\newcommand{\SPD}          {\rm{SPD}\xspace}
\newcommand{\SDD}          {\rm{SDD}\xspace}
\newcommand{\SSD}          {\rm{SSD}\xspace}
\newcommand{\TPC}          {\rm{TPC}\xspace}
\newcommand{\TRD}          {\rm{TRD}\xspace}
\newcommand{\VZERO}        {\rm{V0}\xspace}
\newcommand{\VZEROA}       {\rm{V0A}\xspace}
\newcommand{\VZEROC}       {\rm{V0C}\xspace}
\newcommand{\Vdecay} 	   {\ensuremath{V^{0}}\xspace}

% 4) PARTICLE SPECIES 
\newcommand{\pip}          {\ensuremath{\pi^{+}}\xspace}
\newcommand{\pim}          {\ensuremath{\pi^{-}}\xspace}
\newcommand{\kap}          {\ensuremath{\rm{K}^{+}}\xspace}
\newcommand{\kam}          {\ensuremath{\rm{K}^{-}}\xspace}
\newcommand{\pbar}         {\ensuremath{\rm\overline{p}}\xspace}
\newcommand{\kzero}        {\ensuremath{{\rm K}^{0}_{\rm{S}}}\xspace}
\newcommand{\lmb}          {\ensuremath{\Lambda}\xspace}
\newcommand{\almb}         {\ensuremath{\overline{\Lambda}}\xspace}
\newcommand{\Om}           {\ensuremath{\Omega^-}\xspace}
\newcommand{\Mo}           {\ensuremath{\overline{\Omega}^+}\xspace}
\newcommand{\X}            {\ensuremath{\Xi^-}\xspace}
\newcommand{\Ix}           {\ensuremath{\overline{\Xi}^+}\xspace}
\newcommand{\Xis}          {\ensuremath{\Xi^{\pm}}\xspace}
\newcommand{\Oms}          {\ensuremath{\Omega^{\pm}}\xspace}
\newcommand{\degree}       {\ensuremath{^{\rm o}}\xspace}
\newcommand{\Hc}           {\ensuremath{\mathrm{H_c}}\xspace}
\newcommand{\Dzero}        {\ensuremath{\mathrm{D^0}}\xspace}
\newcommand{\Dplus}        {\ensuremath{\mathrm{D^+}}\xspace}
\newcommand{\Dminus}        {\ensuremath{\mathrm{D^-}}\xspace}

\newcommand{\Dstar}        {\ensuremath{\mathrm{D^{*+}}}\xspace}
\newcommand{\Dstarzero}        {\ensuremath{\mathrm{D^{*0}}}\xspace}
\newcommand{\Ds}           {\ensuremath{\mathrm{D_s^+}}\xspace}
 \newcommand{\Bs}           {\ensuremath{\mathrm{B_s^0}}\xspace}
 \newcommand{\Bplus}           {\ensuremath{\mathrm{B^+}}\xspace}
  \newcommand{\Bzero}           {\ensuremath{\mathrm{B^0}}\xspace}
  \newcommand{\Bc}           {\ensuremath{\mathrm{B^+_c}}\xspace}

\newcommand{\Dsstar}{\ensuremath{\mathrm{D_s^{*+}}}\xspace}

\newcommand{\Dsminus}    
{\ensuremath{\mathrm{D_s^-}}\xspace}
\newcommand{\Lc}           {\ensuremath{\Lambda_\mathrm{c}^+}\xspace}
\newcommand{\Lcgeneric}           {\ensuremath{\Lambda_\mathrm{c}}\xspace}
\newcommand{\Lcminus}           {\ensuremath{\Lambda_\mathrm{c}^-}\xspace}
\newcommand{\LcExcitedOne}{\ensuremath{\Lambda_\mathrm{c}(2595)^+}\xspace}
\newcommand{\LcExcitedTwo}{\ensuremath{\Lambda_\mathrm{c}(2625)^+}\xspace}
\newcommand{\SigmacZero}           {\ensuremath{\Sigma_\mathrm{c}(2455)^{0}}\xspace}
\newcommand{\SigmacZeroPlus}           {\ensuremath{\Sigma_\mathrm{c}(2455)^{0,++}}\xspace}
\newcommand{\SigmacZeroExcited}           {\ensuremath{\Sigma_\mathrm{c}(2520)^{0}}\xspace}

\newcommand{\Sigmac}           {\ensuremath{\Sigma_\mathrm{c}^{0,+,++}}\xspace}
\newcommand{\Sigmacgeneric}           {\ensuremath{\Sigma_\mathrm{c}}\xspace}

\newcommand{\XicZero}      {\ensuremath{\Xi_\mathrm{c}^0}\xspace}
\newcommand{\XicPlus}      {\ensuremath{\Xi_\mathrm{c}^+}\xspace}
\newcommand{\XicPlusZero}  {\ensuremath{\Xi_\mathrm{c}^{0,+}}\xspace}
\newcommand{\Omegac}       {\ensuremath{\Omega_\mathrm{c}^0}\xspace}
\newcommand{\Jpsi}         {\ensuremath{\mathrm{J}/\psi}\xspace}
\newcommand{\PsiTwos}         {\ensuremath{\psi(\mathrm{2S})}\xspace}
\newcommand{\UpsilonOneS}         {\ensuremath{\mathrm{\Upsilon (1S)}}\xspace}
\newcommand{\ccbar}        {\ensuremath{\mathrm{c\overline{c}}}\xspace}
\newcommand{\bbbar}        {\ensuremath{\mathrm{b\overline{b}}}\xspace}

\newcommand{\Xib}          {\ensuremath{\Xi_\mathrm{b}^{0,-}}\xspace}
\newcommand{\DzerotoKpi}   {\ensuremath{\mathrm{D^0\to K^-\pi^+}}}
\newcommand{\Lb}     {\ensuremath{\Lambda_\mathrm{b}^0}\xspace}
\newcommand{\Lbbar}           {\ensuremath{\bar{\Lambda}_\mathrm{b}^0}\xspace}

\newcommand{\lowptbin}{\ensuremath{0<\pt<1}~\GeVc}

\newcommand{\LcD} {\ensuremath{\Lc/\Dzero}\xspace}
\newcommand{\QQbar}           {\ensuremath{\mathrm{Q\bar{Q}}\xspace}}

% 5) MODELS 
\newcommand{\tamu}         {\textsc{tamu}\xspace}
\newcommand{\pythiasix}    {\textsc{pythia6}\xspace}
\newcommand{\herwig}    {\textsc{herwig}\xspace}
\newcommand{\pythiaeight}  {\textsc{pythia8}\xspace}
\newcommand{\pythiaeightprecise}{\textsc{pythia8.243}\xspace}
\newcommand{\pythiasixprecise}{\textsc{pythia6.4.25}\xspace}
\newcommand{\pythia}       {\textsc{pythia}\xspace}
\newcommand{\hijing}       {\textsc{hijing}\xspace}
\newcommand{\hijingprecise}{\textsc{hijing v1.383}\xspace}
\newcommand{\fonll}        {\textsc{fonll}\xspace}
\newcommand{\evtgen}       {\textsc{EvtGen}\xspace}

\title[Article Title]{ MUSIC: A Multi-Purpose Detector Concept for Physics at the 10 TeV Muon Collider }

%%=============================================================%%
%% Prefix	-> \pfx{Dr}
%% GivenName	-> \fnm{Joergen W.}
%% Particle	-> \spfx{van der} -> surname prefix
%% FamilyName	-> \sur{Ploeg}
%% Suffix	-> \sfx{IV}
%% NatureName	-> \tanm{Poet Laureate} -> Title after name
%% Degrees	-> \dgr{MSc, PhD}
%% \author*[1,2]{\pfx{Dr} \fnm{Joergen W.} \spfx{van der} \sur{Ploeg} \sfx{IV} \tanm{Poet Laureate} 
%%                 \dgr{MSc, PhD}}\email{iauthor@gmail.com}
%%=============================================================%%
\author[a]{Paolo Andreetto}
\author[b,c]{Nazar Bartosik}
\author[d]{Andrea Bersani}
\author[e,f]{Daniele Calzolari}
\author[g]{Massimo Casarsa}
\author[h,i]{Vittoria Ludovica Ciccarella}
\author[h]{Elisa Di Meco}
\author[j]{Ruben Gargiulo}
\author[a]{Alessio Gianelle}
\author[g,k]{Carlo Giraldin}
\author[l]{Karol Krizka}
\author[f]{Anton Lechner}
\author[m]{Luigi Longo}
\author[a,e]{Donatella Lucchesi}
\author[a,e]{Leonardo Palombini}
\author[b]{Nadia Pastrone}
\author[h]{Ivano Sarra}
\author[n]{Lorenzo Sestini}
\author[m,o]{Rosamaria Venditti}
\author[a,e]{Davide Zuliani}

\affil[a]{INFN Sezione di Padova, Padua, Italy}
\affil[b]{INFN Sezione di Torino, Turin, Italy}
\affil[c]{Universit\`a del Piemonte Orientale, Vercelli, Italy}
\affil[d]{INFN Sezione di Genova, Genoa, Italy}
\affil[e]{Universit\`a di Padova, Padua, Italy}
\affil[f]{European Organization for Nuclear Research, Geneva, Switzerland}
\affil[g]{INFN Sezione di Trieste, Trieste, Italy}
\affil[h]{INFN Laboratori Nazionali di Frascati, Frascati, Italy}
\affil[i]{Sapienza Universit\`a di Roma, Rome, Italy}
\affil[j]{INFN Sezione di Roma 1, Roma, Italy}
\affil[k]{Universit\`a di Trieste, Trieste, Italy}
\affil[l]{University of Birmingham, Birmingham, United Kingdom}
\affil[m]{INFN Sezione di Bari, Bari, Italy}
\affil[n]{INFN Sezione di Firenze, Florence, Italy}
\affil[o]{Universit\`a di Bari, Bari, Italy}
\affil[]{\newline \newline \newline \small Contact: Davide Zuliani (\href{davide.zuliani@cern.ch}{davide.zuliani@cern.ch})
}

%%==================================%%
%% sample for unstructured abstract %%
%%==================================%%

\abstract{\unboldmath
%A proof-of-concept of MUSIC, a multi-purpose detector conceived for high-precision and ultra–high-energy physics studies in the challenging environment of $\sqrt{s}=10$ TeV muon-antimuon collisions, is presented.
This work presents a proof of concept for MUSIC, a multi-purpose detector conceived for high-precision and ultra–high-energy physics studies in the challenging environment of $\sqrt{s}=10$ TeV muon–antimuon collisions.
The detector features a central tracking system, electromagnetic and hadronic calorimeters, and dedicated muon detectors. This paper outlines the main design elements of each subdetector, with an emphasis on the effects of machine-induced backgrounds and the reconstruction strategies employed for key physics objects. Performance results for electrons, photons, muons, and jets are reported, and studies of jet flavour identification are discussed.}

\maketitle
\clearpage

\section{Detector design for $\mathbf{\sqrt{s}=10}$ TeV muon collisions}
\label{sec:physics}
Designing a proof-of-concept detector for muon–an\-ti\-mu\-on collisions at a center-of-mass energy of 10~TeV is an unprecedented challenge; leptonic collisions at such an energy scale became conceivable only with the recent consolidation of muon collider studies~\cite{epjc}. The guiding principles underlying our design are summarized in Ref.~\cite{annual}.

This concept builds on state-of-the-art detector technologies, while assuming incremental improvements that are realistically achievable by the time a full technical design is undertaken. By then, further advances may also enable entirely new detector concepts. The detector model presented here serves as a benchmark for evaluating the achievable performance in the reconstruction of physics objects under the challenging experimental conditions of a multi-TeV muon collider.

Among today's highest-priority topics in particle physics are the high-precision determination of the Higgs couplings to bosons and fermions and the measurement of the Higgs potential parameters. At $\sqrt{s}=10$ TeV, muon-antimuon collisions produce a copious number of Higgs bosons, which offers a unique opportunity for precision studies. A suitably designed detector could achieve precision Higgs coupling measurements that surpass those of any other proposed future-collider experiment. 

Studies in Ref.~\cite{annual} indicate that, at $\sqrt{s}=10$ TeV, the transverse momentum components of the Higgs boson decay products are comparable to those observed at $\sqrt{s}=3$ TeV. In contrast, their angular distribution becomes increasingly forward-peaked, with many decay products emerging at small polar angles relative to the beam axis. This motivates improved detector coverage in the forward region.

At the same time, searches for physics beyond the Standard Model require robust detector redundancy for central, very high-momentum objects, such as multi-TeV leptons, and the capability for high-precision measurements. Concretely, this entails multiple, independent momentum determinations to control resolution and charge misidentification at the multi-TeV scale; nearly hermetic, finely segmented calorimetry to provide precise missing momentum (or energy); extended rapidity coverage and precision timing to suppress machine-induced backgrounds and resolve displaced activity; and reconstruction optimized for unconventional signatures such as disappearing or emerging tracks, displaced vertices, and non-standard jets.

Considerable effort is currently devoted to developing detector concepts for future muon colliders~\cite{maia}, aiming to meet the unprecedented experimental challenges at multi-TeV energies. Within this context, this work presents the current MUSIC (MUon System for Interesting Collisions) detector concept, outlining the design and performance of each sub-detector. Although this is a proof-of-concept study, it demonstrates that machine-induced backgrounds can be effectively controlled, enabling performance that is competitive with, and in some cases superior to, that projected for other future-collider experiments at more advanced stages of development.

\section{The MUSIC detector}
\label{sec:detector}

\begin{figure*}[t]
    \centering
    \includegraphics[width=0.7\linewidth]{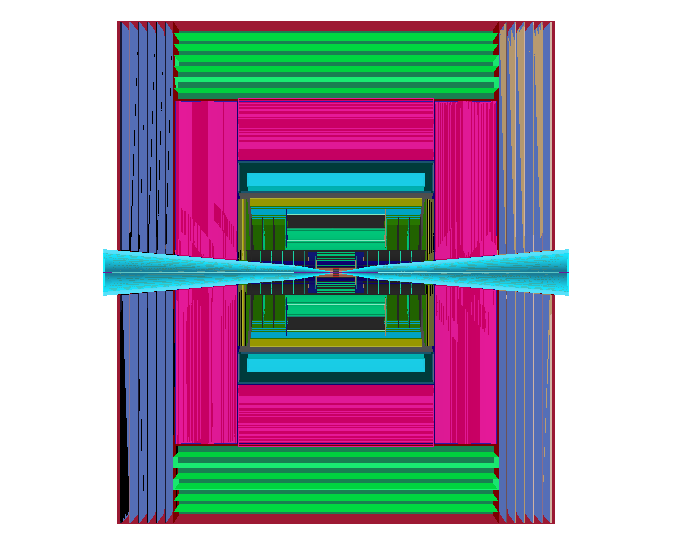}
    \caption{Layout of the MUSIC detector in the $y$-$z$ plane.}
    \label{fig:MUSIC}
\end{figure*}

The principal design constraint for a detector intended to operate at $\sqrt{s}=10$ TeV muon-antimuon collisions is the expected rate and composition of machine-induced backgrounds. 
The various background contributions at $\sqrt{s}=1.5$ and $3$ TeV due to muon decays (beam-induced background, BIB) have been documented in Refs.~\cite{epjc, annual, 3tevphysics}. Preliminary studies in Ref.~\cite{ichep24MDI10} indicate that at $\sqrt{s}=10$ TeV, incoherent $e^+e^-$ pair production (IPP) also becomes an important source of machine-induced background, particularly affecting subdetectors closest to the beamline, such as the tracking system. This has mainly driven the choice of the solenoidal magnetic-field strength.
The key features of the machine-induced backgrounds relevant to the detector design and their impact on physics object reconstruction are summarized below. 
The interaction region (IR) layout corresponds to the lattice version 0.8~\cite{lattice_EU24}, which was frozen 
for the 2026 update studies of the European Strategy for Particle Physics
%for the European Strategy for Particle Physics 2025 studies 
and is referred to as the “EU24” configuration.

\medskip\noindent
The effects of particles originating from the decay of muons in the beams are mitigated by inserting two cone-shaped tungsten shields, referred to as ``nozzles'', along the beamline within the detector, between the interaction point (IP) and the detector extremities. 
In Fig.~\ref{fig:MUSIC}, where the detector concept is illustrated in the $y$-$z$ plane\footnote{A right-handed reference system is used in MUSIC, with the origin at the center of the detector, the nominal collision point: the $z$-axis is aligned with the direction of the clockwise-circulating $\mu^+$ beam, the $y$ axis points upward, and the $x$ axis lies on the plane of the collider ring.}, the nozzles are clearly visible in cyan.

Interactions of high-momentum electrons, positrons, and photons with the shielding generate showers that are not fully contained by the nozzles. The resulting leakage produces a low-energy, high-multiplicity background flux that reaches the detector. Its impact differs across subsystems, the tracking system, the electromagnetic and hadronic calorimeters, and the muon spectrometer, because their timing acceptances, energy thresholds, and material responses are distinct. Background effects and mitigation strategies are therefore discussed within each sub-detector dedicated section.

\noindent
Incoherent $e^+e^-$ pairs are produced through collisions of real or virtual photons emitted by muons in the counter-rotating bunches. Their production has been studied using a sample of $e^+e^-$ pairs generated at the IP with the GUINEA-PIG code~\cite{guinea-pig}, and the resulting particles are subsequently transported through the IR geometry using FLUKA~\cite{fluka}.
Because a sizeable fraction originates near the IP, these pairs are not intercepted by the shielding nozzles and can reach the inner tracking volume. However, their trajectories are shaped by the solenoidal magnetic field; as shown in Ref.~\cite{ichep24MDI10}, a 5~T field confines a significant fraction of them close to the beamline, thereby reducing occupancy in the central tracker.
This remains a partial mitigation; residual rates and energy deposits must be assessed together with the muon-decay–induced background, and their combined impact quantified for each sub-detector.

\begin{figure}[ht]
    \centering
    \includegraphics[width=1\linewidth]{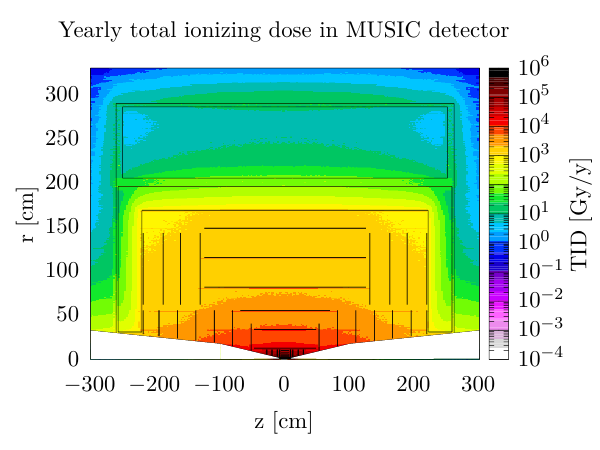}
    \caption{Total ionizing dose per year from Ref.~\cite{ESPPUimc}, as defined in the text. 
    \label{fig:TID}}
\end{figure}

\begin{figure}[ht]
    \centering
    \includegraphics[width=1\linewidth]{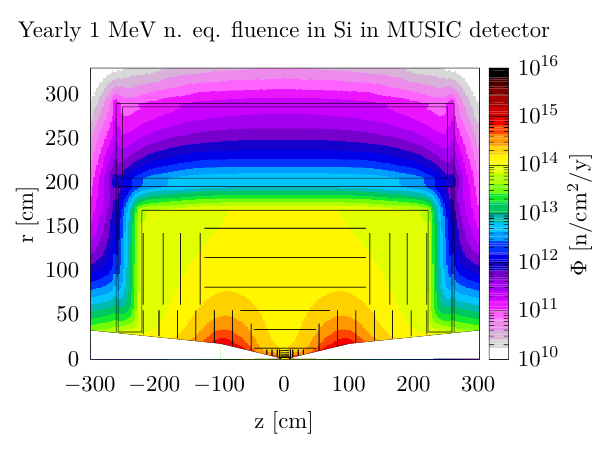}
    \caption{1-MeV neutron-equivalent fluence in silicon per year from Ref.~\cite{ESPPUimc}, as defined in the text.
    \label{fig:neq}}
\end{figure}

Based on these studies, the detector’s radiation environment is expected to be dominated by particles originating from muon beam decays. 
The total ionizing dose (TID), which quantifies ionization damage in the detectors' materials, reaches its maximum near the vertex detector up to approximately 1 MGy/year, as shown in Fig.~\ref{fig:TID}, assuming 139 operational days.
The 1-MeV neutron-equivalent fluence, used to characterize displacement damage in silicon sensors, peaks in the inner tracker due to neutron leakage from the shielding nozzles, reaching values of approximately $10^{15}$\,n/cm$^{2}$/year, as shown in Fig.~\ref{fig:neq}.
These levels inform the choice of radiation-hard sensor technologies, cooling and power distribution, and further optimization of the shielding and layout.
\subsection{Superconducting solenoid magnet}
\label{sec:magnet}
For precise measurements of particle momenta, a magnetic field of 5~T is foreseen in the tracker volume. In principle, no magnetic field is required in electromagnetic and hadronic calorimeters, and even in the muon system; therefore, field value and uniformity are being optimised only in the central volume of the detector. Presently, a main solenoid between the electromagnetic and hadronic calorimeter is foreseen: the clearance for the solenoid cryostat is 5~m in length, 4~m in diameter, and 80~cm in thickness in radial direction. In this study, we assume a uniform magnetic field of 5 T parallel to the $z$ axis inside the tracker, while in the HCAL, a constant field of 1.8~T oriented in the opposite direction is considered.
If compared to operating magnets, the MUSIC solenoid is fairly similar to the CMS one~\cite{CMS}, which features a central field of 4~T in a larger volume\footnote{Niobium Titanium alloy can operate with significantly higher magnetic fields at 4.2~K, the expected operating temperature of the MUSIC magnet.}. The cable technology from the CMS solenoid has been taken as the basis for the MUSIC cable development, which has demonstrated reliable use with similar forces and stresses over 10,000 hours of operation. Presently, no company routinely produces aluminium stabilised superconducting cable, but a joint effort between industries, CERN, and KEK is resuming this technology, therefore suggesting that the MUSIC magnet will be feasible within the foreseen schedule for the accelerator complex.

\begin{figure}[ht!]
    \centering
   \includegraphics[width=0.8\linewidth]{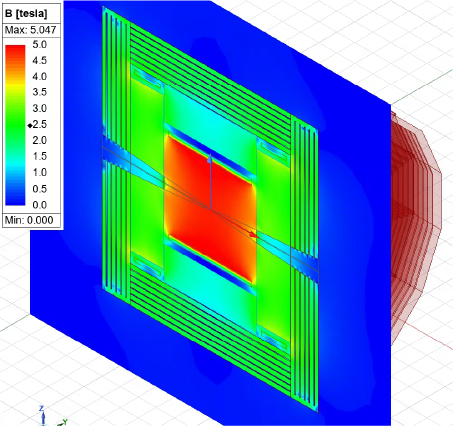}
    \caption{Magnetic field map on a vertical cross-section of the MUSIC detector}
   \label{fig:fie}
\end{figure}

A very preliminary design for the MUSIC solenoid has been developed by a part of the INFN group involved in the design and construction of the CMS solenoid. This design is still under development and will be optimised in the future, in parallel with the development of the detector design. 
Presently, the magnet design features two additional coils, out of the hadronic calorimeter end-caps, inside the instrumented iron yoke. With the same overall current density as CMS (14~$\textrm{A}/\textrm{mm}^2$), four layers of cable in the main solenoid and three layers on the lateral ones, a central field of 4.75~T can be achieved. The introduction of lateral coils allows for a significant increase in central field ($0.5$~T) and in field uniformity, which is better than $\pm10\%$ with the three coils and slightly above $\pm12\%$ with the central coil only. A field map on a cross section of the detector is shown in Fig.~\ref{fig:fie}: only coils, iron yoke, electromagnetic calorimeter, and nozzles have been modelled.

\subsection{Tracking system}
\begin{figure}[ht!]
    \centering
    \includegraphics[width=1.0\linewidth]{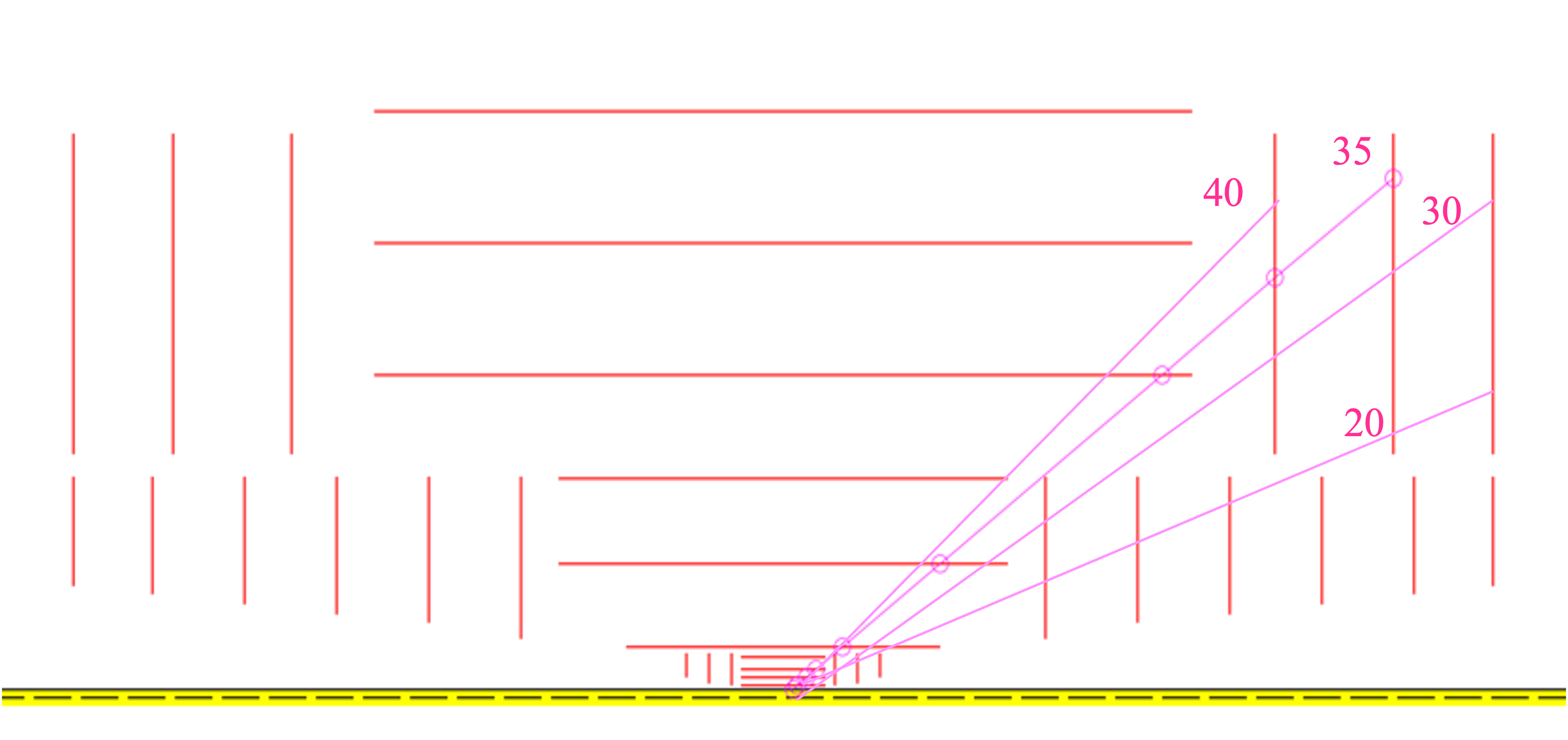}
    \caption{Schematic view of the tracking system in the $r$-$z$ plane. Different angles showing the angular acceptance of the tracking barrel and endcaps are shown with purple lines.}
    \label{fig:tracker}
\end{figure}

\begin{figure}[ht!]
    \centering
    \includegraphics[width=1.0\linewidth]{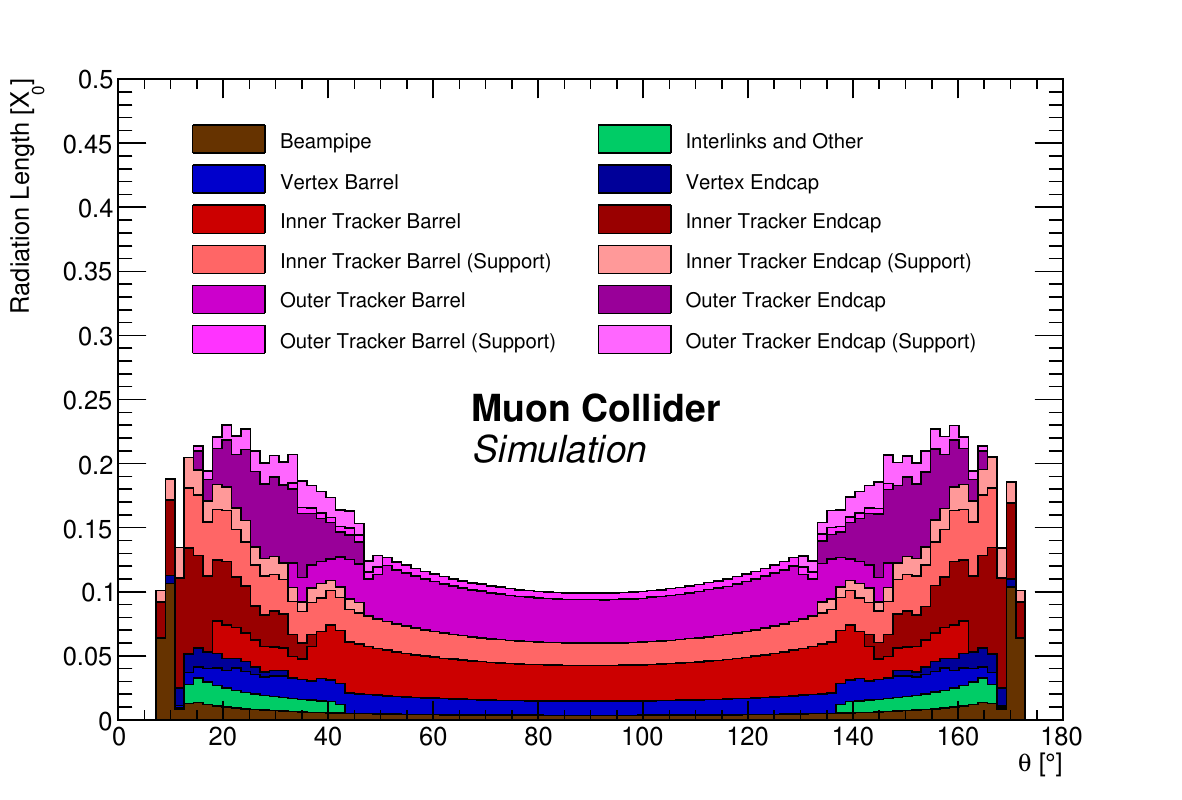}
    \caption{Radiation length $X_0$ of the tracking system as a function of the polar angle $\theta$.}
    \label{fig:mm}
\end{figure}

The tracking system is optimized to maximize geometric acceptance and to provide redundant measurements that mitigate background-induced inefficiencies. With respect to the  $\sqrt{s}=3$ TeV case~\cite{3tevphysics}, the longitudinal extent of the silicon barrel layers has been increased to enhance the coverage at large pseudorapidities, while retaining the barrel-plus-disks layout. The radial and $z$
positions of successive layers—both barrels and disks—are chosen to eliminate geometric cracks, with intentional overlaps in the most critical regions to stabilize pattern recognition, seeding, and momentum determination. A schematic view of the tracking system in the $r$-$z$ plane is shown in Fig.~\ref{fig:tracker}, while Fig.~\ref{fig:mm} shows the fraction of radiation length $X_0$ of the whole tracker as a function of $\cos{\theta}$ of the incoming particle, where $\theta$ is the polar angle.

\noindent
The tracking system consists of three sub-detectors:
\medskip
\begin{description}
    \item[Vertex Detector (VXD):] It is the innermost system, consisting of five cylindrical barrel layers, each 26.0 cm long, positioned at radii ranging from 2.9 to 10.1 cm from the beam axis. The forward and backward regions feature four endcap disks, oriented transverse to the beamline and located at distances of $|z| = 18.0$ to 36.6 cm from the interaction point.
    A layout of $25 \times 25$ $\mu$m$^2$ pixels is assumed, with a hit spatial resolution of 5 $\mu$m $\times$ 5 $\mu$m and a hit time resolution of $30$ ps. 
    \item[Inner Tracker (IT):] It consists of 50 $\mu$m $\times$ 1 mm macropixel modules arranged in three barrel layers, at radii from 16.4 to 55.4 cm, and seven disks on either side at $|z|$ from 60.4 to 219.0 cm. The first two barrel layers are 96.32 cm long, while the third measures 138.46 cm. Hit spatial and time resolutions of 7 $\mu$m $\times$ 90 $\mu$m and 60 ps are assumed, respectively.     
    \item[Outer Tracker (OT):] It has three 252.8-cm long barrel layers at radii between 81.9 and 148.6 cm and four endcap disks at $|z|$ from 141.0 to 219.0 cm. It features 50 $\mu$m $\times$ 1 mm macropixel with a hit spatial resolution of 7 $\mu$m $\times$ 90 $\mu$m and a hit time resolution of 60 ps.
\end{description}

\subsection{Calorimeters}
The MUSIC detector foresees an electromagnetic and a hadronic calorimeter. The electromagnetic calorimeter is specifically designed for muon-antimuon collisions, while the hadronic calorimeter is adapted from the CLIC Collaboration detector developed for $e^+e^-$ collisions at $\sqrt{s}=3$ TeV~\cite{CLICDET}.
%\medskip
\begin{description}
    \item[Electromagnetic calorimeter (ECAL):] The ECAL is a semi-homogeneous electromagnetic crystal cal\-o\-rim\-e\-ter with longitudinal segmentation (CRILIN)~\cite{crilin}. It consists of $1 \times 1 \times 4$-cm$^3$ lead-fluorite crystals arranged in six layers for a total of 26.5 radiation lengths. The cylindrical barrel section has an inner radius of 169.0 cm and is 442.0 cm long. The endcaps, shaped as disks, have inner and outer radii of 31.0 cm and 196.0 cm, respectively, and are positioned at $|z| =$ 230.7 cm.  It operates inside the 5~T magnetic field.
    \item[Hadronic calorimeter (HCAL):] The HCAL is an iron-scintillator sampling calorimeter composed of 70 layers of 2-cm thick iron absorbers and $3 \times 3$ cm$^2$ scintillator pads, each 0.3 cm thick, totaling approximately seven nuclear interaction lengths. It consists of a central cylindrical part measuring 501.8 cm in length and 290.2 cm in radius, along with two endcaps positioned at $|z| = 257.9$ cm. The endcaps have inner and outer radii of 32.0 cm and 475.6 cm, respectively. Positioned outside the superconducting solenoid, the HCAL iron absorber also functions as a return yoke for the magnetic field flux. Alternative technologies mostly focused on MPGD-HCAL are currently under study~\cite{Longo:2024pk}.
\end{description}

\subsection{Muon system}
The final technology for the muon detectors has not yet been selected. 
The current MUSIC simulation is modeled on the resistive-plate chambers employed by the CLIC detector~\cite{CLICDET}.
%The current detector simulation uses what is in the original CLIC detector~\cite{CLICDET}, resistive-plate chambers. 
It features seven barrel layers, each measuring 888.8 cm in length, with radii ranging from 480.6 cm to 680.0 cm, and six endcap layers at $|z|$ positions between 444.4 cm and 590.0 cm, with inner and outer radii of 49.3 cm and 680.0 cm, respectively. 
%A hit spatial resolution of 1 cm $\times$ 1 cm and a hit time resolution $\sigma_t = 100$ ps are assumed.
A cell size of 3 cm $\times$ 3 cm and a hit time resolution of $\sigma_t = 100$ ps are assumed. Alternative technologies are under study.
%
%%%%%%%%%%%%%%%%%%%%%%%%%
%
\section{ Software pipeline description}
\label{sec:software}
\begin{figure}
    \centering
    \includegraphics[width=1.0\linewidth]{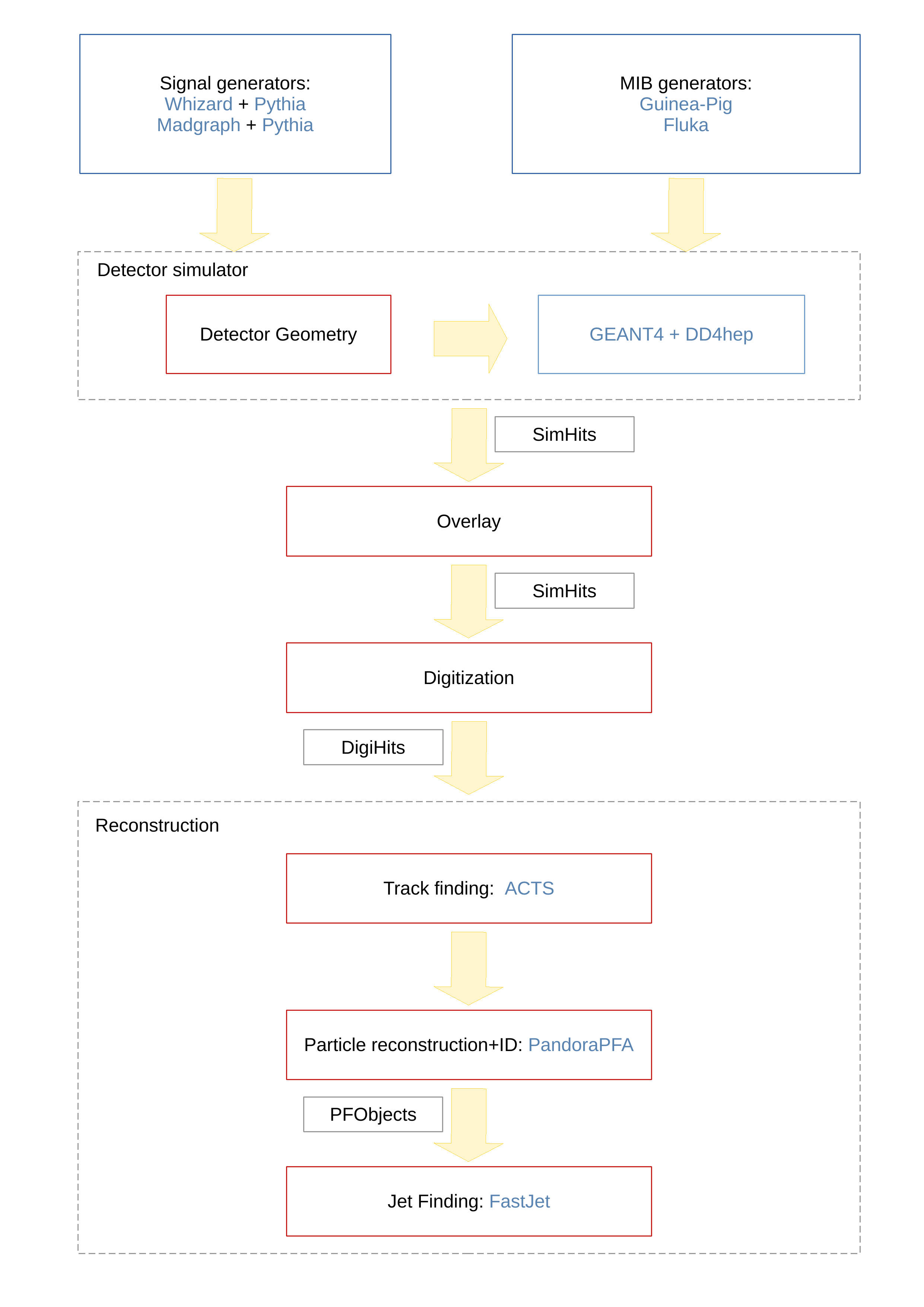}
    \caption{Software pipeline, red boxes are components developed or integrated by the Muon Collider Group}
    \label{fig:swpipeline}
\end{figure}

The end-to-end software chain, depicted in Fig.~\ref{fig:swpipeline}, from the generation of signal and machine-induced-background (MIB) events to physics-object reconstruction, combines widely used HEP packages with components developed and tuned specifically for the high-energy muon collider environment. Ref.~\cite{MuCSoftware} provides a detailed description of the framework; here, we present only a brief summary to aid in the interpretation of the detector-performance results.
\begin{description}
    \item[Generation:] Signal and physics–background events are generated with \textsc{Whizard}~\cite{whizard} and/or \textsc{MadGraph}~\cite{madgraph}, in combination with \textsc{PYTHIA}~\cite{ref:pythia8}, as a Monte Carlo event generator. Machine-induced backgrounds are simulated with \textsc{FLUKA}~\cite{fluka} and \textsc{GUINEA-PIG}~\cite{guinea-pig} and transported through the IR to the detector. To study the performance of the detector, particle guns and PYTHIA are used.
    \item[Detector simulation:] Particle transport and interactions in the passive and active materials of the detector are simulated with \textsc{Geant4}~\cite{ref:geant4} and \textsc{DD4hep}~\cite{DD4hep}. The core engine of \textsc{Geant4} provides the main functionalities for a full simulation, while \textsc{DDh4ep}, built on top of it, simplifies the designing of the detector geometry through a description-oriented approach. The outcome of the simulation consists on a collection of hit-related parameters - like position, time, and energy deposit - called \texttt{SimHit}. The \texttt{SimHit}, like other information produced in the rest of the pipeline, is structured according to standard data models, required for guaranteeing the interoperability among different software stacks. The data models supported by the Muon Collider framework are \textsc{LCIO}~\cite{LCIO} and \textsc{EDM4hep}~\cite{EDM4hep}.
    \item[Overlay:] A key step in event processing is the overlay of machine-induced background hits with signal and physics background hits. This is performed by merging the \texttt{SimHits} collections to produce a single combined output.
   \item[Digitization:] Subdetector-specific digitization emulators convert \texttt{SimHits} into readout channels and are tuned to mitigate machine-induced backgrounds at the data-collection level (e.g., timing windows, energy thresholds, zero suppression). The outcome of the digitization process is a collection of hit parameters called \texttt{DigiHit}.
    \item[Reconstruction:] Subdetector primitives are reconstructed and then combined in a global reconstruction to form physics objects (tracks, vertices, leptons, jets,
    missing energy,
    %\(E_{\mathrm{T}}^{\text{miss}}\), 
    etc.). The track reconstruction is performed via the ACTS toolkit \cite{acts}; the framework implements a``seeding and track followin'' approach by which each track is built point by point, starting from an initial seed, leveraging the prediction of a combinatorial Kalman filter. Selecting the correct seed is crucial to obtain good quality tracks and is strongly related to the geometry of the detector and the physical processes. The reconstruction of single-particle objects (muons, photons, electrons) is performed via the PandoraPFA toolkit \cite{pandora}, using a specifically-optimized version of the LCContent library originally designed for linear collider studies. PandoraPFA exploits a Particle Flow reconstruction approach, where calorimetric energy deposits are clustered and matched individually to tracks, forming single-particle candidates. The detailed description of the sub-algorithms, performing the clustering of calorimeter hits and the reconstruction of electrons and muons, is provided in the respective Sections (\ref{sec:calo}, \ref{sec:muon}). The jet clustering algorithm is $k_T$ \cite{kt}, taking the reconstructed single-particle candidates as input. An efficient implementation of the algorithm is provided by \textsc{FastJet}~\cite{ref:fastjet}. The secondary vertex finding exploits the LCFIPlus software \cite{lcfiplus}, which takes the reconstructed tracks as inputs and produces a set of vertex candidates, associating tracks based on their compatibility with a common origin point. 

\end{description}

%
%%%%%%%%%%%%%%%%%%%%%%%%%
%
\section{Physics objects reconstruction and identification performance}
\label{sec:performance}
The results reported here were obtained with the software pipeline described in Sec.~\ref{sec:software}. For each physics object, the principal reconstruction procedures and selection requirements are outlined and the resulting performance is discussed. The machine-induced backgrounds from muon decay and incoherent $e^+e^-$ production are always included.

\subsection{Track reconstruction}
\label{sec:tracking}
\begin{figure}[ht]
    \centering
    \includegraphics[width=1\linewidth]{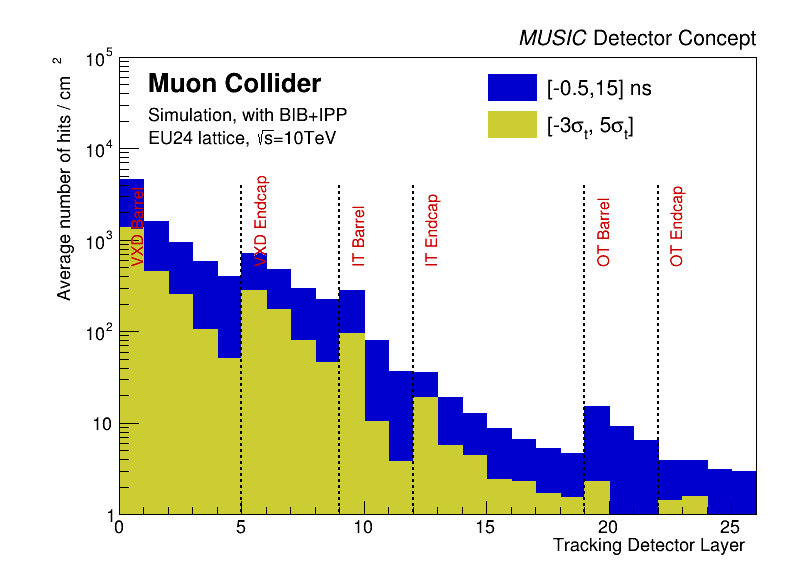}
    \caption{Average hit occupancy in the tracking system layers due to BIB and IPP under two different arrival time requirements on the sensor. $\sigma_t$ is 30 ps (60 ps) for the VXD (IT, OT).}
    \label{fig:occupancy}
    \includegraphics[width=1\linewidth]{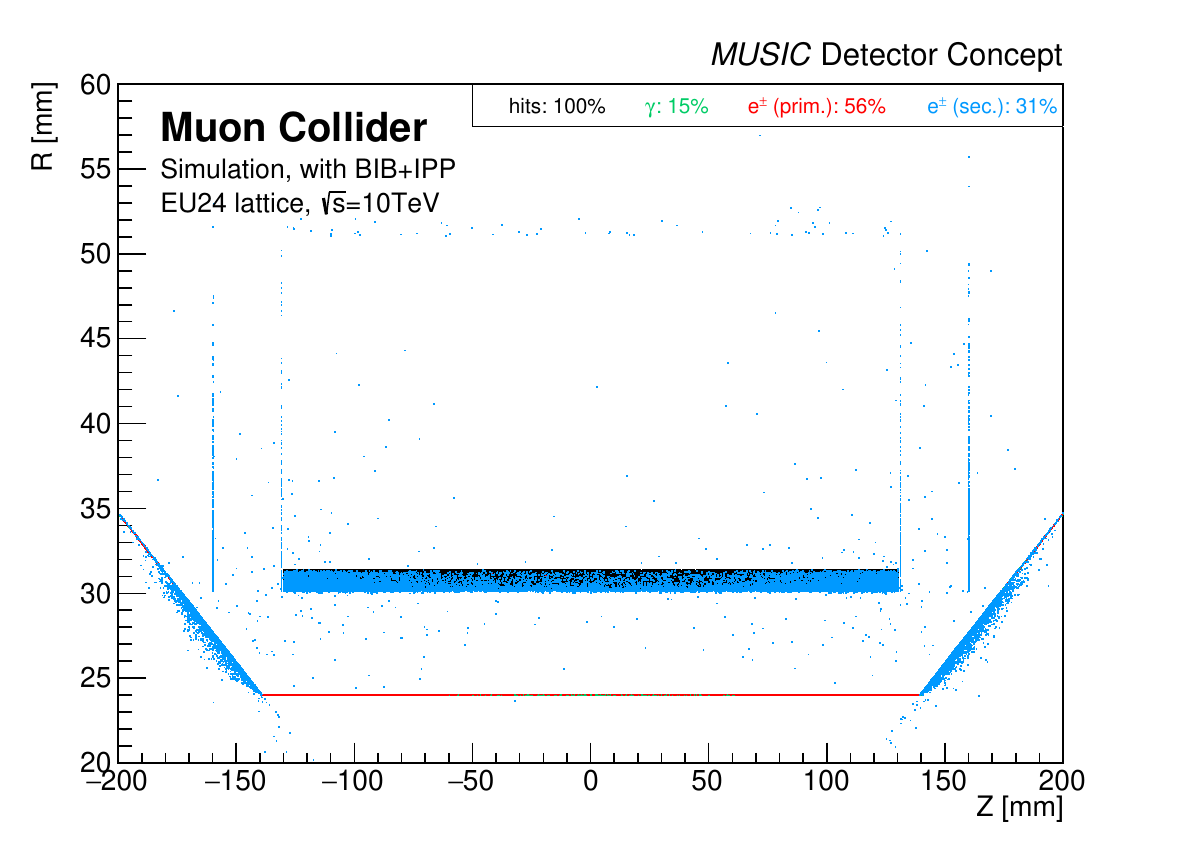}
    \caption{Origin of the beam-induced background particles in $R$-$z$ plane creating hits in the first layer of VXD. Green points show primary photons, while red and blue points describe primary and secondary photons, respectively.}
    \label{fig:BIB_origin_VXD}
    
\end{figure}

Track reconstruction is among the most demanding tasks at high-energy muon colliders, primarily because of the large flux of spurious hits from machine-induced backgrounds. This is illustrated in Fig.~\ref{fig:occupancy}, which shows the occupancy of each tracker layer, defined as the average number of hits per cm$^2$. The vertex detector (VXD) is the most affected subsystem, owing to its proximity to the interaction point and small radii, necessitating dedicated reconstruction strategies to include it in the track definition. 
Future optimizations of the machine-detector interface and VXD will be guided by studies of the origin and species of particles producing hits in the inner tracker. Figure~\ref{fig:BIB_origin_VXD} shows, for hits recorded in the first VXD layer, the distribution of the production point in the $R$-$z$ plane, of the particle generating them. Only one beam, impinging from right to left, is considered.
The dominant contribution (red) arises from primary electrons/positrons emerging directly from the nozzles, whose outline is clearly visible in the figure.
The second-largest contribution (blue) consists of secondary electrons/positrons, produced when primaries interact with the forward and backward silicon disks and with the second VXD layer; an additional component originates from particles entering from the opposite side nozzle.
A small fraction is due to photons.
The reconstruction proceeds with track finding and fitting, followed by a final selection that retains only high-purity tracks. The ACTS~\cite{acts} program constructs 3D space-points from raw tracker hits via simple clustering of adjacent channels; these space-points are then used for seeding. To control combinatorics under high occupancies, seeding is restricted to hits in the second, third, and fourth VXD layers, deliberately excluding the innermost layer at this stage. In the subsequent fit, however, hits from the first VXD layer are incorporated to restore optimal impact-parameter resolution. 
The present work does not include the optimization of a primary-vertex finding algorithm, which should account for the predicted machine transverse and longitudinal beam-spot sizes of $\sigma_{x,y} = 1\,\mu\text{m}$ and $\sigma_{z} = 1.5$\,mm. Hence, each study assumes a primary-vertex position of $(0,0,0)$ \cite{annual}. The final selection suppresses fake tracks and duplicates using standard criteria.
\begin{figure}[ht!]
    \centering

    \includegraphics[width=1\linewidth]{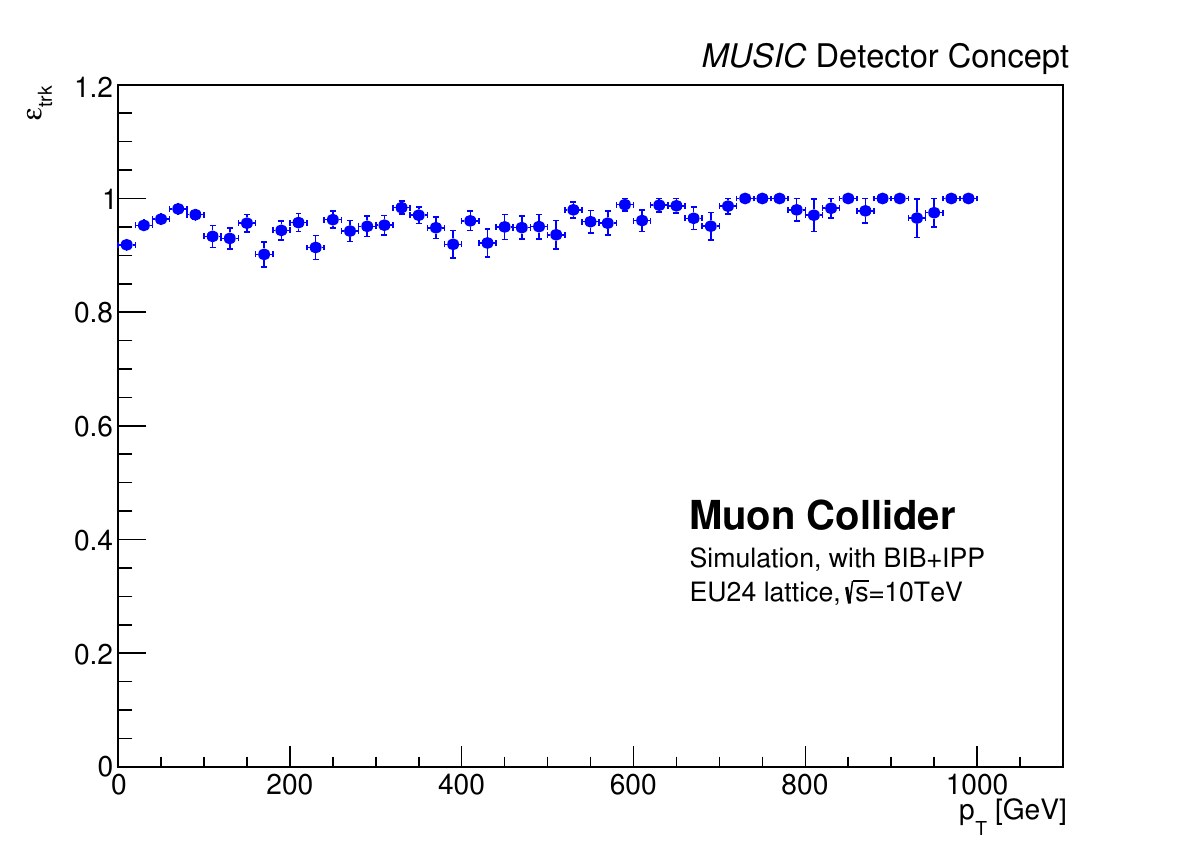}
    \caption{Track reconstruction efficiency as a function of the muon transverse momentum.}
    \label{fig:track_eff}

    \includegraphics[width=1\linewidth]{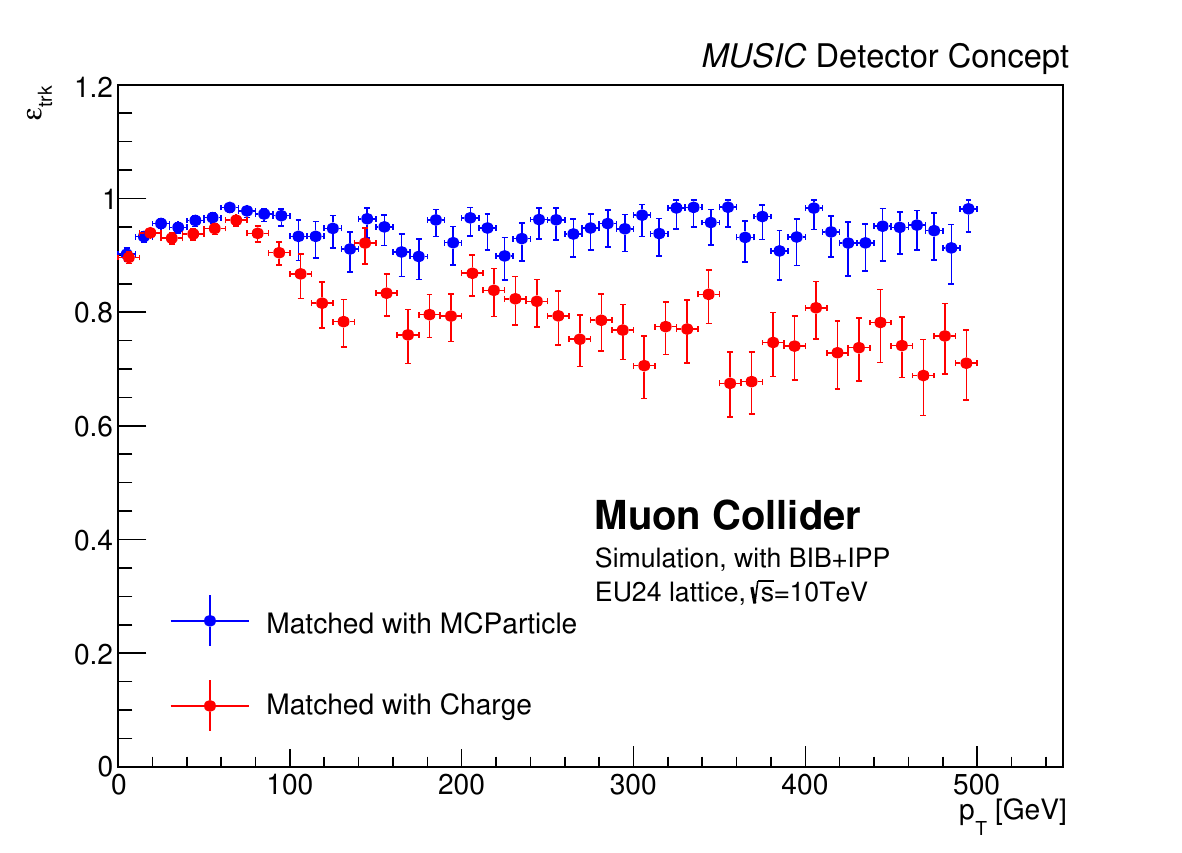}
    \caption{Track reconstruction efficiency with the correct curvature sign (red points) compared to the overall track reconstruction efficiency (blue points) as a function of the muon transverse momentum.}
    \label{fig:track_eff_charge}
\end{figure}
Tracking performance is evaluated using a sample of single muons produced at the IP and generated flat in momentum, azimuthal angle, and polar angle; no muon identification is applied. Due to the high-occupancy environment typical of a muon collider, it is not possible to distinguish between fake tracks and tracks generated by the machine-induced background. To this extent, tight requirements on the total number of hits used in the track reconstruction are imposed in order to suppress machine-induced background. These requirements are optimised to drastically reduce the number of fake tracks per event, while retaining a good reconstruction efficiency for real tracks. Figure~\ref{fig:track_eff} shows the track reconstruction efficiency $\varepsilon_{\mathrm{trk}}$ for muons as a function of the muon transverse momentum. The reconstruction efficiency $\varepsilon_{\mathrm{trk}}$ is defined as $\varepsilon_{\mathrm{trk}}=N^{\mathrm{reco}}_{\mathrm{tracks}}/N^{\mathrm{true}}_{\mathrm{tracks}}$, where $N^{\mathrm{reco}}_{\mathrm{tracks}}$ is the number of reconstructed tracks that are matched to a true track\footnote{A reconstructed track is said to be ``truth-matched" to a true track if $\Delta R<0.005$.}, while $N^{\mathrm{true}}_{\mathrm{tracks}}$ is the number of true simulated tracks. Despite the tight requirements imposed to suppress machine-induced background, the efficiency remains high across the full range of $p_T$. Because high-energy muon collisions yield very high-momentum tracks, the probability of assigning the wrong curvature, therefore the wrong charge sign, to a track is non-negligible. Figure~\ref{fig:track_eff_charge} compares, as a function of the muon $p_T$, the efficiency for reconstructing tracks with the correct curvature sign (red points) to the inclusive reconstruction efficiency. At the highest momenta, the charge-misidentification fraction reaches values of the order of 15\%.
\begin{figure}[ht]
    \centering

    \includegraphics[width=1\linewidth]{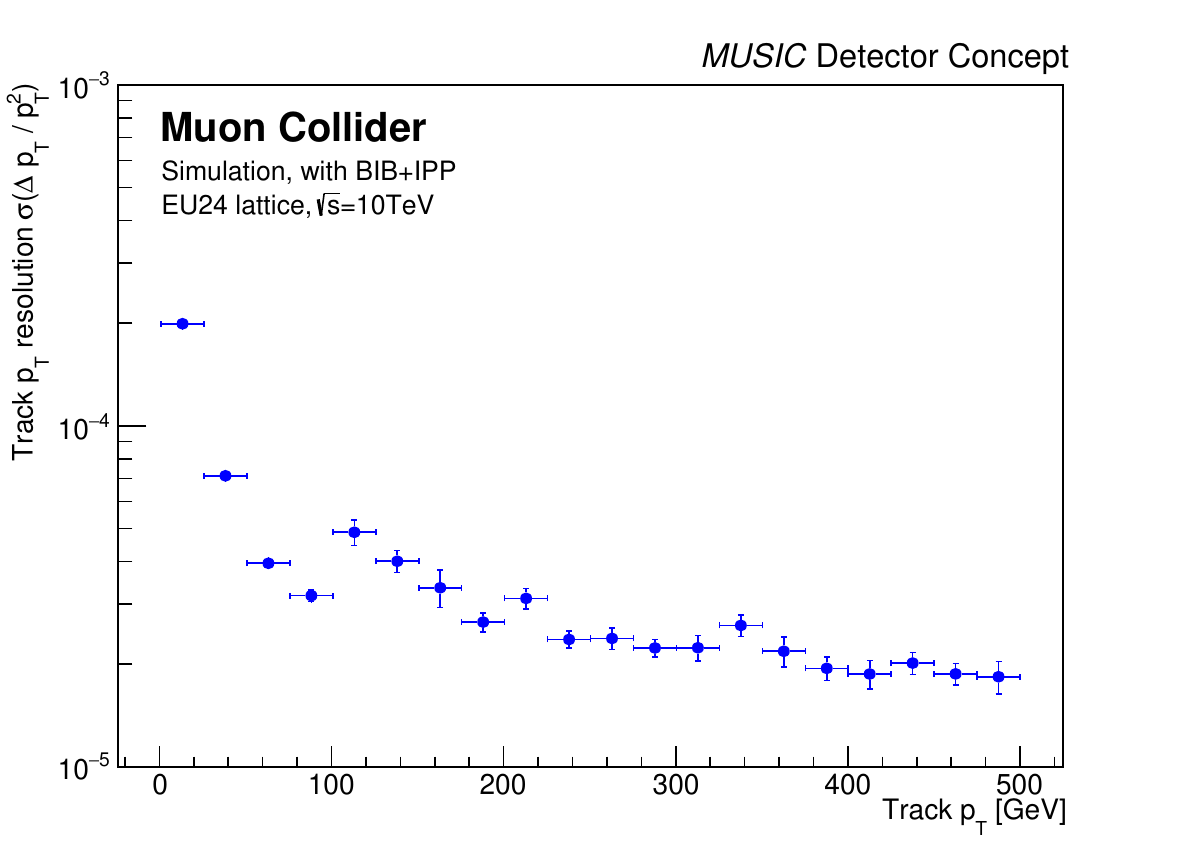}
    \caption{Transverse momentum resolution $\sigma(\Delta p_{T}/p^2_{T})$ as a function of the muon transverse momentum.}
    \label{fig:track_ptreso}

    \includegraphics[width=1\linewidth]{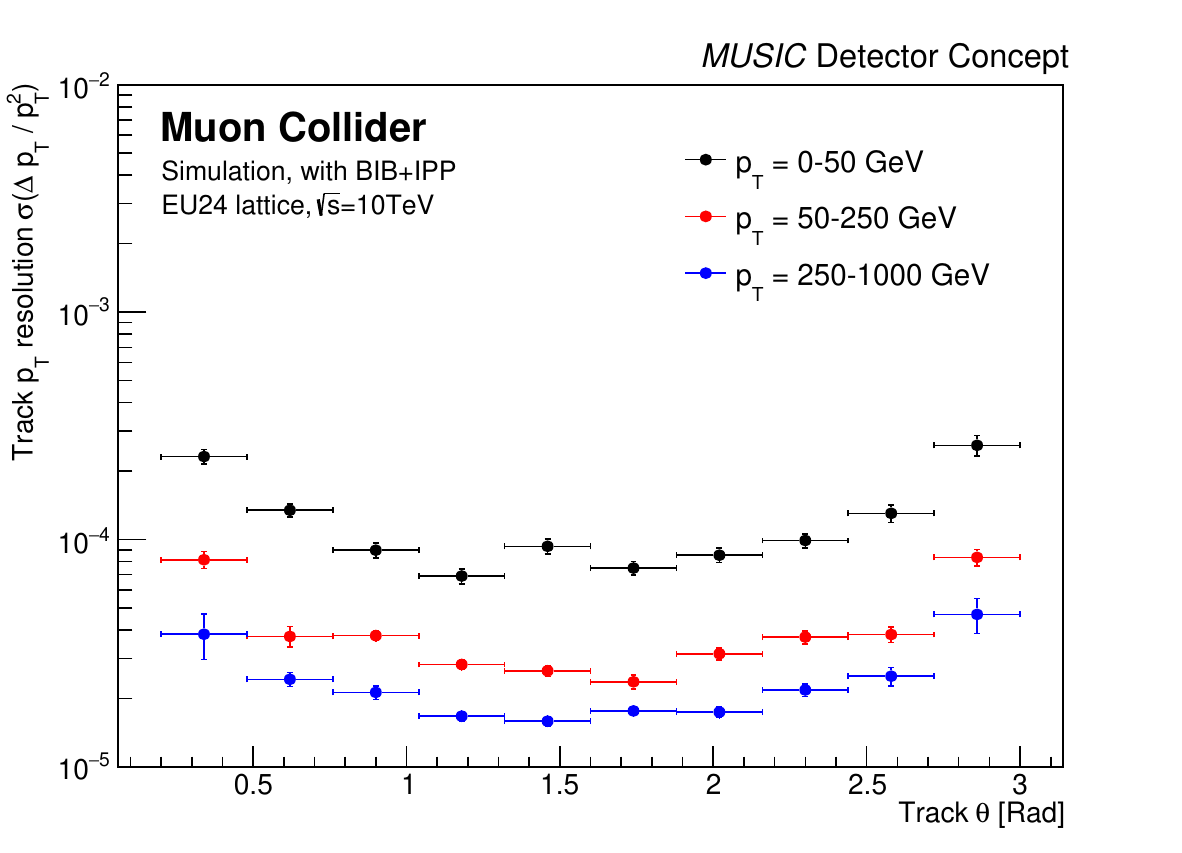}
    \caption{Transverse momentum resolution $\sigma(\Delta p_{T}/p^2_{T})$ as a function of the muon polar angle for different muon $p_{\mathrm{T}}$. }
    \label{fig:track_anglereso}
\end{figure}
High performance is evidenced not only by the reconstruction efficiency but also by the achieved resolution on the track parameters. For example, Fig.~\ref{fig:track_ptreso} shows the transverse momentum resolution as a function of the muon transverse momentum $p_T$. The transverse momentum resolution $\sigma(\Delta p_T/p_T^2)$ is defined as the standard deviation of the $(p_{T,\mathrm{reco}}-p_{T,\mathrm{true}})/p_{T,\mathrm{true}}^2$ distribution, where $p_{T,\mathrm{true}}$ ($p_{T,\mathrm{reco}}$) is the transverse momentum of the true (reconstructed) muon. In the crowded muon-collider environment, tracks with momentum in the range [0 - 50] GeV exiting at $90^\circ$ have a transverse momentum resolution of order $ 10^{-4}$, comparable to those obtained under much cleaner conditions like $e^+e^-$~\cite{fcc-feasibility}. 
The resolution also depends on the track polar angle, $\theta$, because of the non-uniform distribution of machine-induced backgrounds and the presence of the shielding nozzles.
As shown in Fig.~\ref{fig:track_anglereso}, a modest resolution degradation is observed for low momentum particles in the forward and backward regions. 

\subsection{Photon and electron identification}
\label{sec:calo}
\begin{figure}[ht]
    \centering
    \includegraphics[width=1\linewidth]{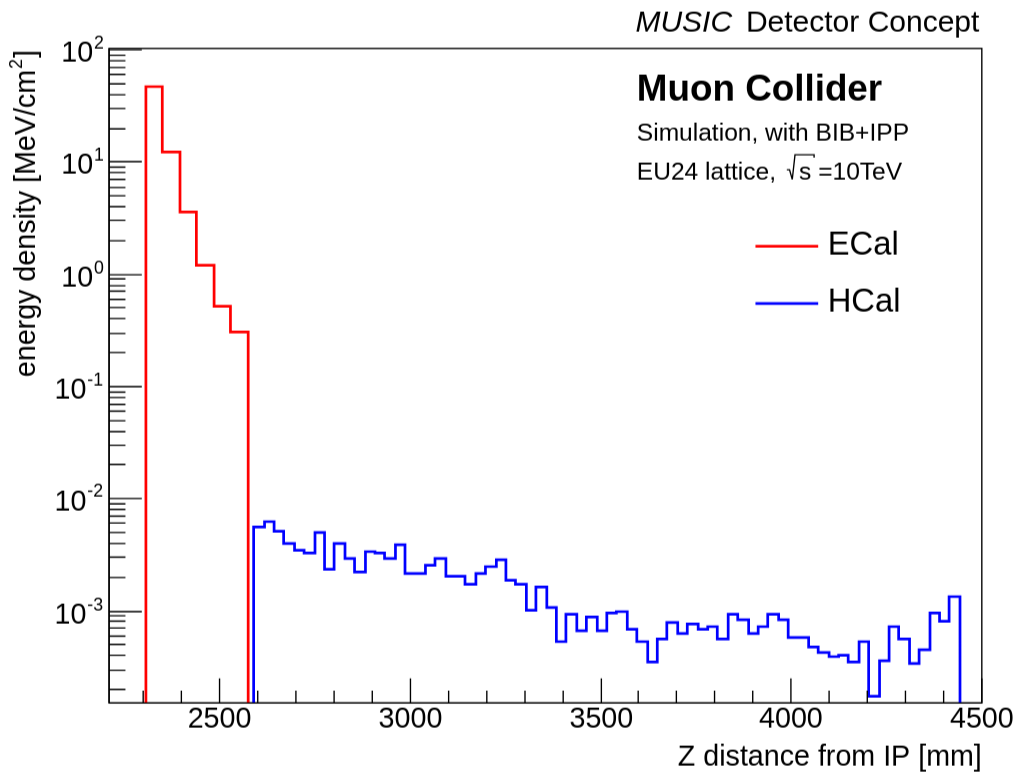}
    \caption{Occupancy of the ECAL and HCAL endcap sections, as a function of the longitudinal distance from the IP.}
    \label{fig:Calo_occupancy}
     \includegraphics[width=1\linewidth]{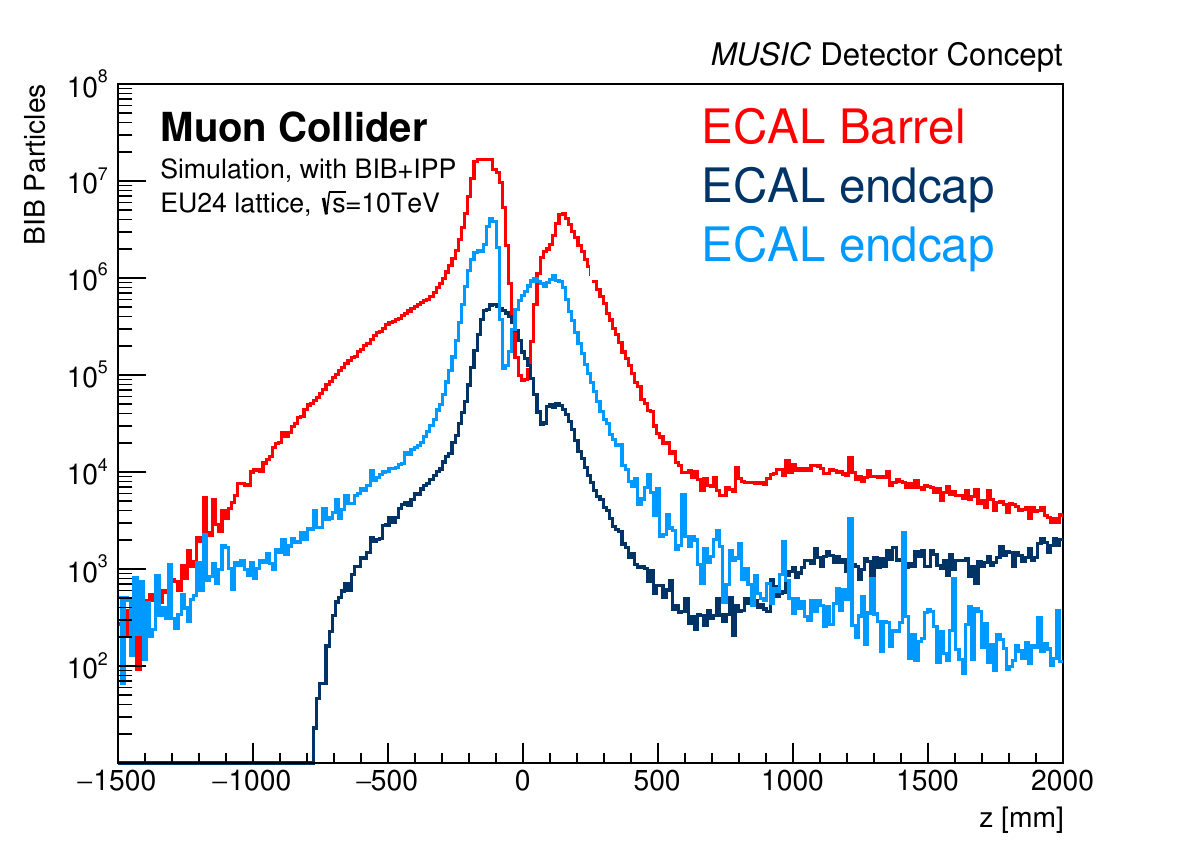}
    \caption{Origin of beam-induced background particles creating hits in the electromagnetic calorimeter. Only one beam is considered, coming from the right.}
    \label{fig:BIB_Calo_origin}
\end{figure}

Despite being located far from the beamline, the ECAL is subjected to substantial background from photons and neutrons originating directly from the nozzles. As shown in Fig.~\ref{fig:Calo_occupancy}, the deposited-energy density per cm$^2$ is very high in the first three layers and decreases rapidly with increasing distance from the interaction point, leaving the HCAL largely unaffected by this background.

As for the VXD, the origins of particles responsible for ECAL hits have been studied using a single muon beam incident from the right. Figure~\ref{fig:BIB_Calo_origin} shows the $z$ distribution separated by region: barrel (red), endcap on the same side as the incoming beam (light-blue), and endcap on the opposite side (dark-blue). In all three regions, a sizeable contribution arises from particles backscattered off the nozzle on the side opposite to the beam. These backscattered particles reach the calorimeter with a significant delay relative to the bunch crossing, providing a timing handle for background suppression.

A dedicated ECAL digitizer has been developed to suppress machine-induced backgrounds. For each layer and subregion, a per-channel threshold is defined as a function of hit arrival time and deposited energy. The subregion granularity, six equal segments along 
$z$ in the barrel and three equally wide annuli in the endcaps, has been optimized using background-only distributions in the corresponding regions. Only deposits exceeding the local threshold are retained. Subsequently, an optimized version of the PandoraPFA clustering algorithm is applied to reconstruct topological clusters for particle-energy determination.
ECAL energies are calibrated in bins of polar angle ($\theta$) and the generator level energy  $E_{MC}$. For each bin, $ E_{MC} / E_{meas}$ is formed,  where $E_{meas}$ is the reconstructed cluster energy. The bin-wise calibration factor $k$ is taken as the average of this distribution. The corrected energy is then $E_{corr}=kE_{meas}$.  This procedure compensates for angular and energy-dependent response non-uniformities.

\begin{figure}
    \centering
    \includegraphics[width=1\linewidth]{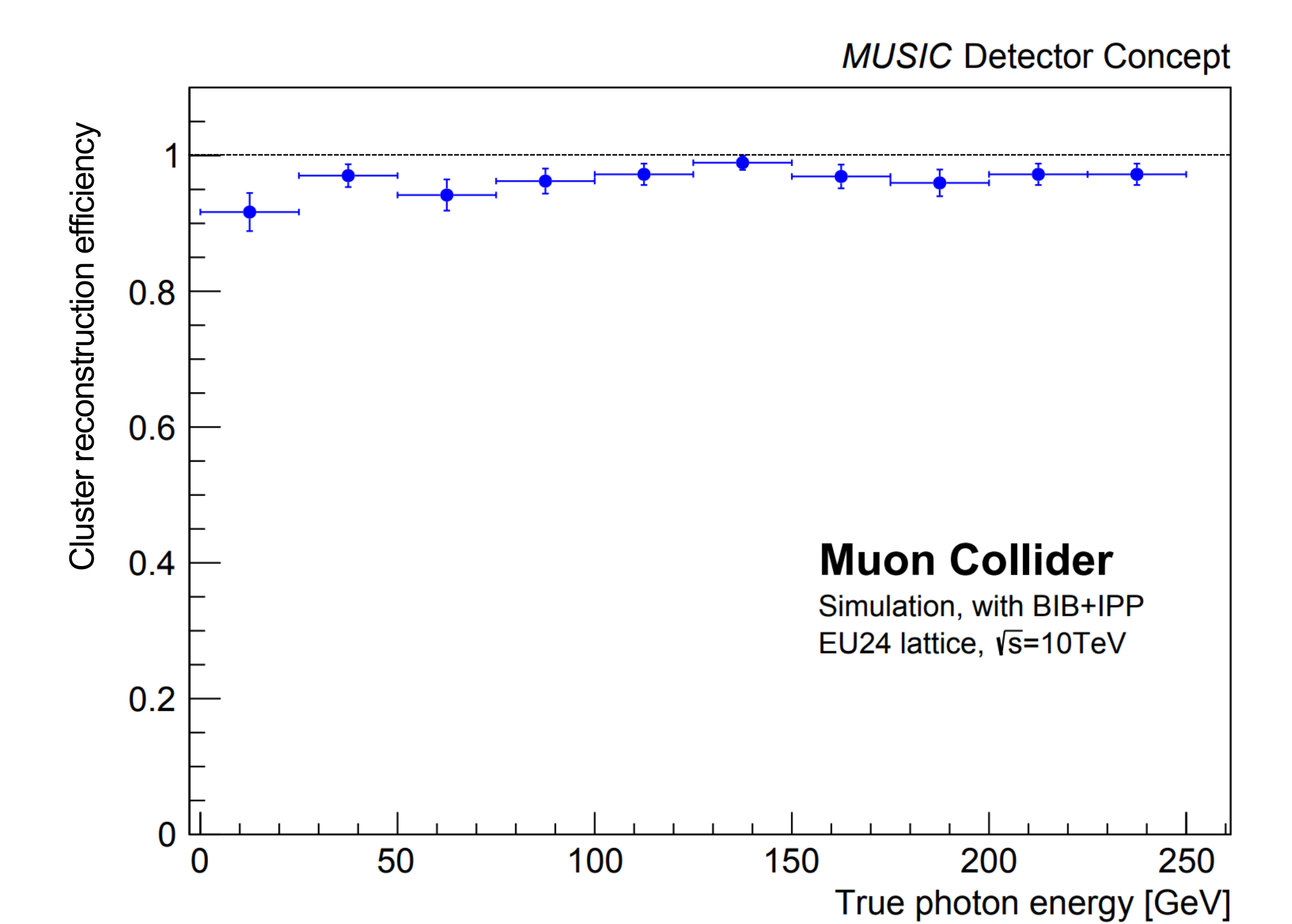}
    \caption{Reconstruction efficiency for photon clusters as a function of Monte Carlo photon energy.}
    \label{fig:cluster_eff}
\end{figure}

The clustering algorithm yields an efficiency $\gtrsim 95\%$, for clusters generated by single photons with true energy $E_{MC}>10$ GeV, as shown in Figure \ref{fig:cluster_eff}

%A photon candidate is defined as an ECAL cluster without associated tracks. The ECAL cluster reconstruction efficiency $\varepsilon$ is evaluated in bins of generator-level (``true'') photon energy $ E_{MC}$ as $\varepsilon=N_{reco}/N_{tot}$. In each $ E_{MC}$ bin, $N_{tot}$ is the number of generated photons within the acceptance, and $N_{reco}$ counts those for which a ECAL cluster is found in a region of $\Delta R = \sqrt{\Delta\theta^2+\Delta\phi^2}<R_0$ from the direction of the true photon, with $R_0 = 0.05$ for the ECAL barrel and $R_0 = 0.04$ for the endcap. If more than one ECAL cluster is found in $\Delta R<R_0$, the one with the smallest $R$ is chosen.The efficiency remains close to 100\% for $20^\circ < \theta < 160^\circ$, while it decreases to approximately 60\% at the acceptance limits $\theta\simeq 10^\circ$ and $170^\circ$, as shown in Fig.~\ref{fig:photon_eff}.

A single procedure is exploited to reconstruct and identify electrons and photons. The energy cluster with their energy-weighted centroid inside ECAL are selected as electromagnetic cluster candidates (\textit{ECAL clusters}). Only clusters with polar angle in the range $13^\circ < \theta < 167^\circ$ are selected, to avoid the high-background region in the immediate vicinity of the nozzles. Tracks are reconstructed as described in Sec.~\ref{sec:tracking} and further spurious-tracks removal procedures are specifically applied at the particle identification level to suppress the contamination from beam-induced background combinatorial tracks.

\begin{figure}[ht]
    \centering
    \includegraphics[width=1\linewidth]{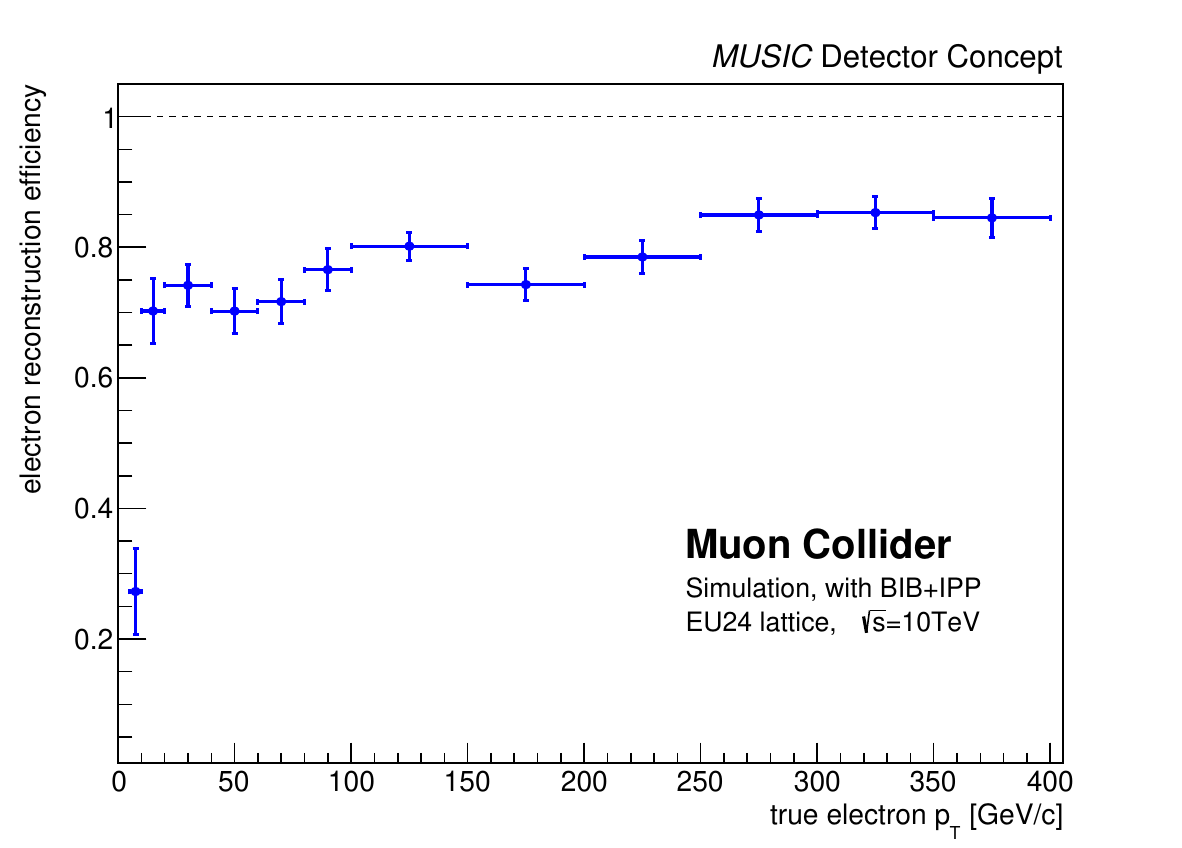}
    \caption{Electron reconstruction efficiency as a function of Monte Carlo electron $p_T$.}
    \label{fig:electron_eff}
    \includegraphics[width=1\linewidth]{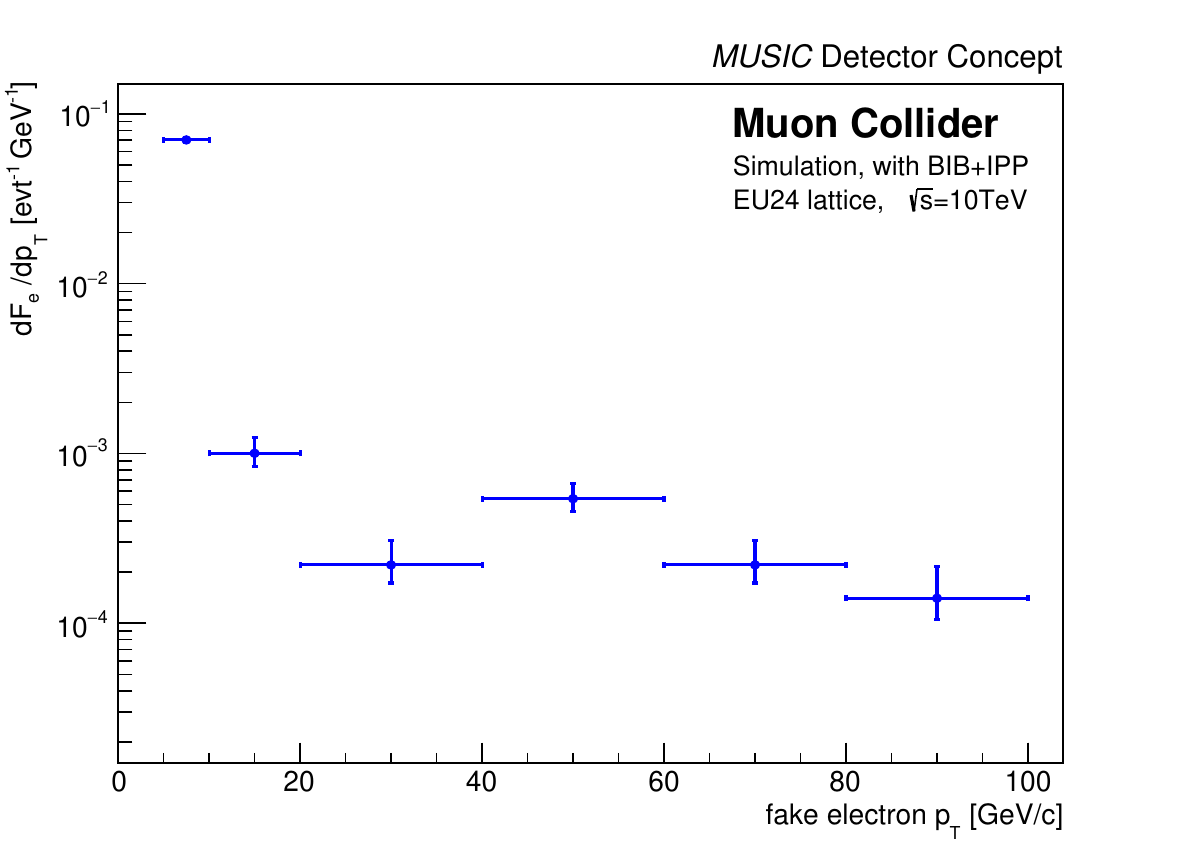}
    \caption{Electron fake rate, as a function of the fake electron's $p_T$.}
    \label{fig:electron_fake}
\end{figure}

An electron is identified as a track with $p_T > 5$ GeV/c, associated to an ECAL cluster with $E_{cl} > 10$ GeV, and it is required that $|E_{cl}-p_{trk}|/p_{trk} < 0.3$ for candidates with $p_T < 40$ GeV/c. These requirements are needed to reduce the probability of random track-cluster associations. The resulting electron reconstruction and identification efficiency is stable between $\sim 75\%$ and $\sim 85\%$ in the full true $p_T$, as in Figure \ref{fig:electron_eff}. The rate of fake-reconstructed electrons due to machine-induced bacgrounds is $\sim 0.4/$evt$^{-1}$, and can be reduced down to $<0.04/$evt$^{-1}$ by requiring a $p_T>10$ GeV/c (Figure \ref{fig:electron_fake}).

\begin{figure}[ht]
    \centering
    \includegraphics[width=1\linewidth]{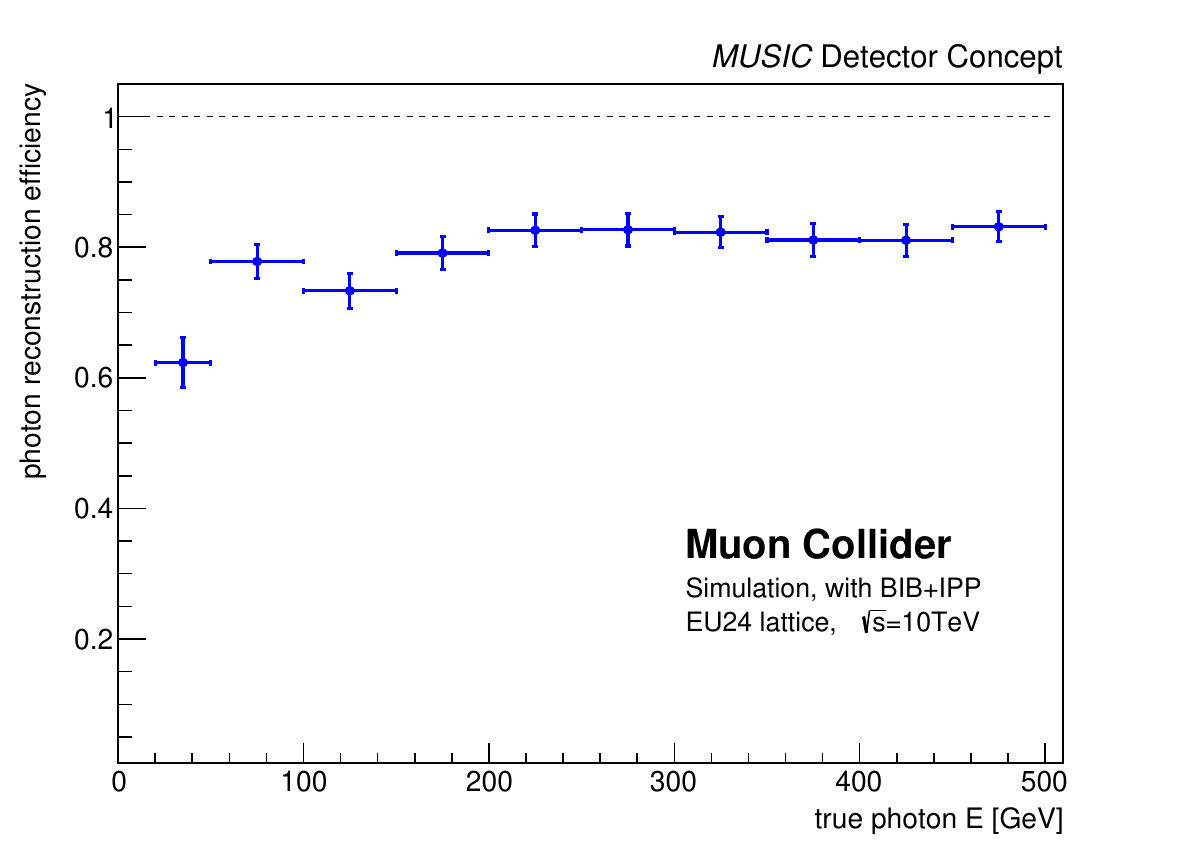}
    \caption{Photon reconstruction efficiency as a function of Monte Carlo photon energy.}
    \label{fig:photon_eff}
    \includegraphics[width=1\linewidth]{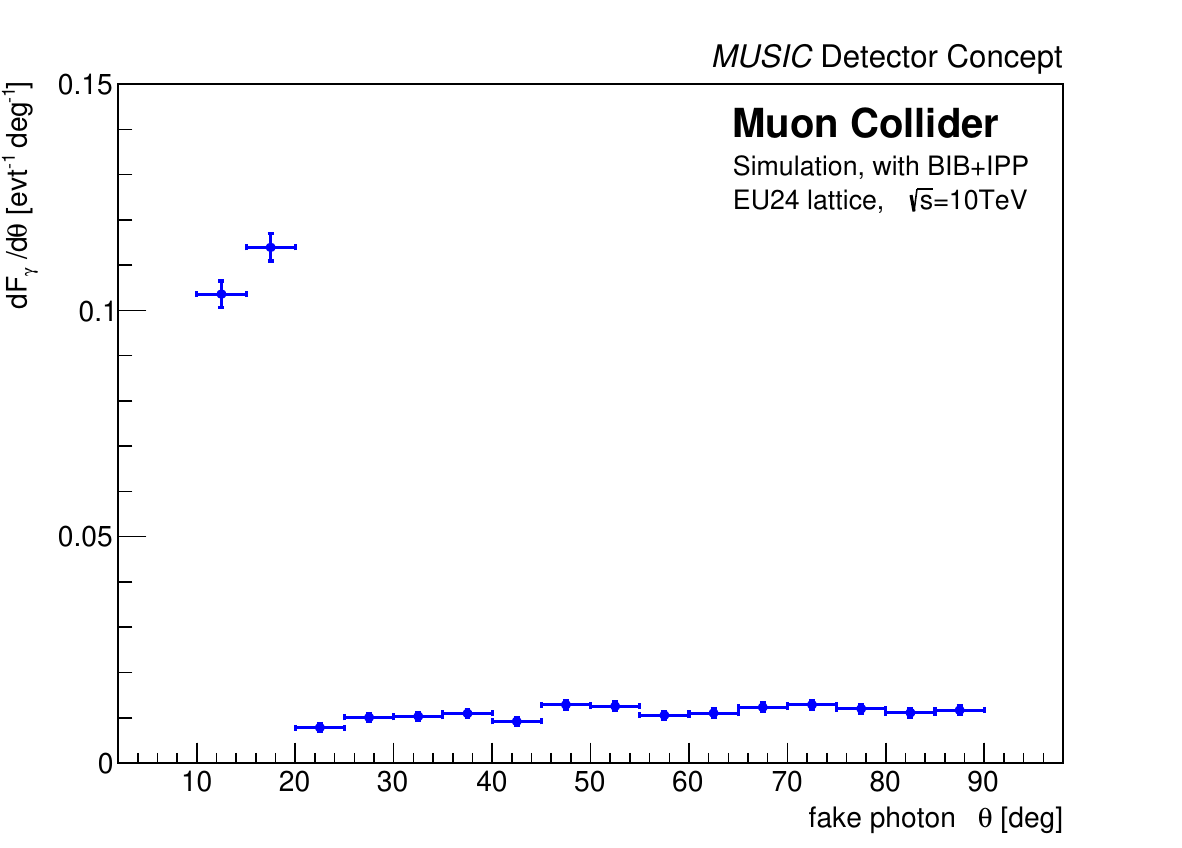}
    \caption{Photons fake rate, as a function of the fake photon's polar angle.}
    \label{fig:photon_fake}
\end{figure}

A photon is identified as an ECAL cluster with $E_{cl} > 20$ GeV, which has not been associated to a track with the aforementioned procedure. The resulting photon reconstruction and identification efficiency is $\sim 80\%$ in the tested true energy range, as in Figure \ref{fig:photon_eff}, being stable at $\sim 95\%$ in the ECAL barrel region and smoothly decreasing to $\sim 40\%$ at small polar angles. The rate of fake-recontructed photons due to machine-induced bacgrounds is $\sim 1.8/$evt$^{-1}$, and it is strongly peaked at small polar angles, due to the proximity of the nozzles (Figure \ref{fig:photon_fake}).

Inefficiencies of the tracking procedure and of the track-cluster association algorithm may cause part of the electrons to be identified as photons, due to their similar energy deposition pattern in ECAL. Vice-versa, accidental track-cluster matches due to the large track density may result in a fraction of photons being identified as electrons. All algorithms can be tuned to obtain different levels of efficiencies, fake rates, and mis-identification probabilities, depending on the analysis requirements. Here, an example working point is presented.

\begin{figure}[ht]
    \centering
    \includegraphics[width=1\linewidth]{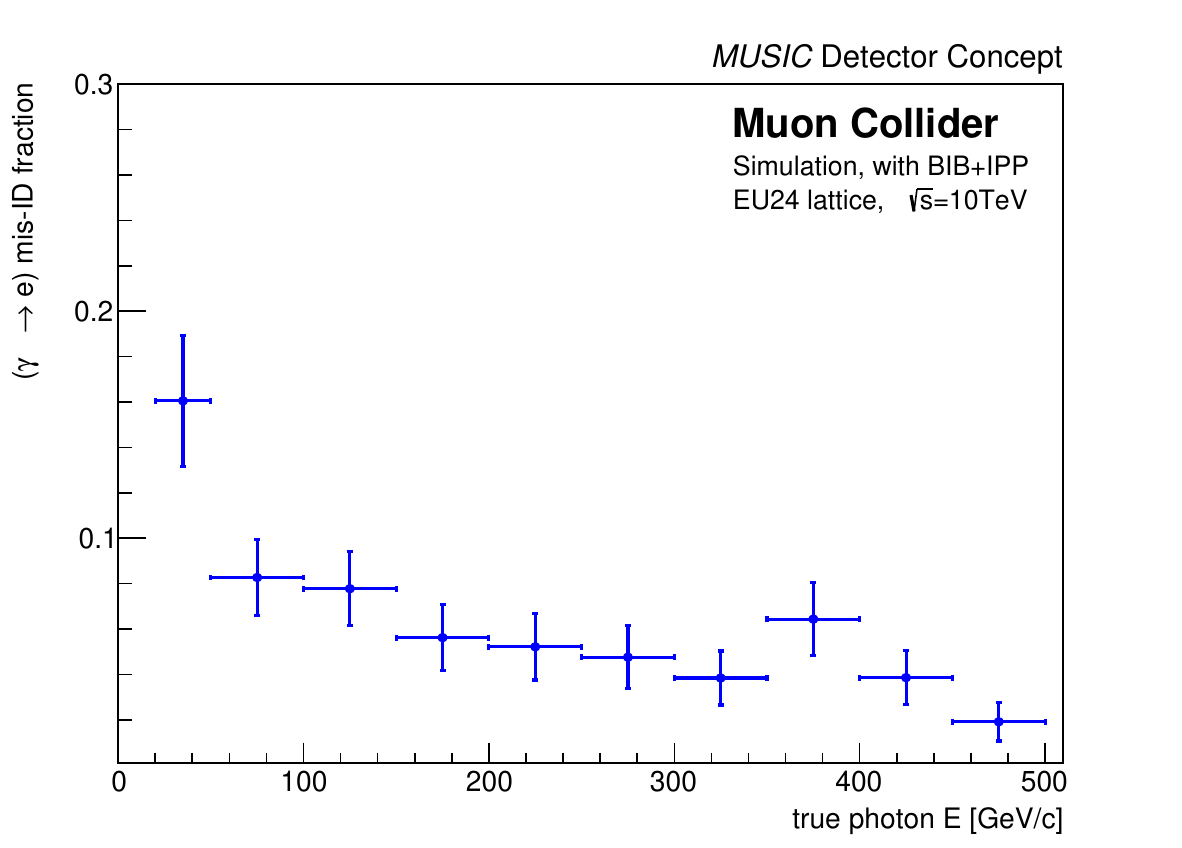}
    \caption{Photons mis-identification probability to electrons, as a function of Monte Carlo photon energy.}
    \label{fig:photon_misid}
    \includegraphics[width=1\linewidth]{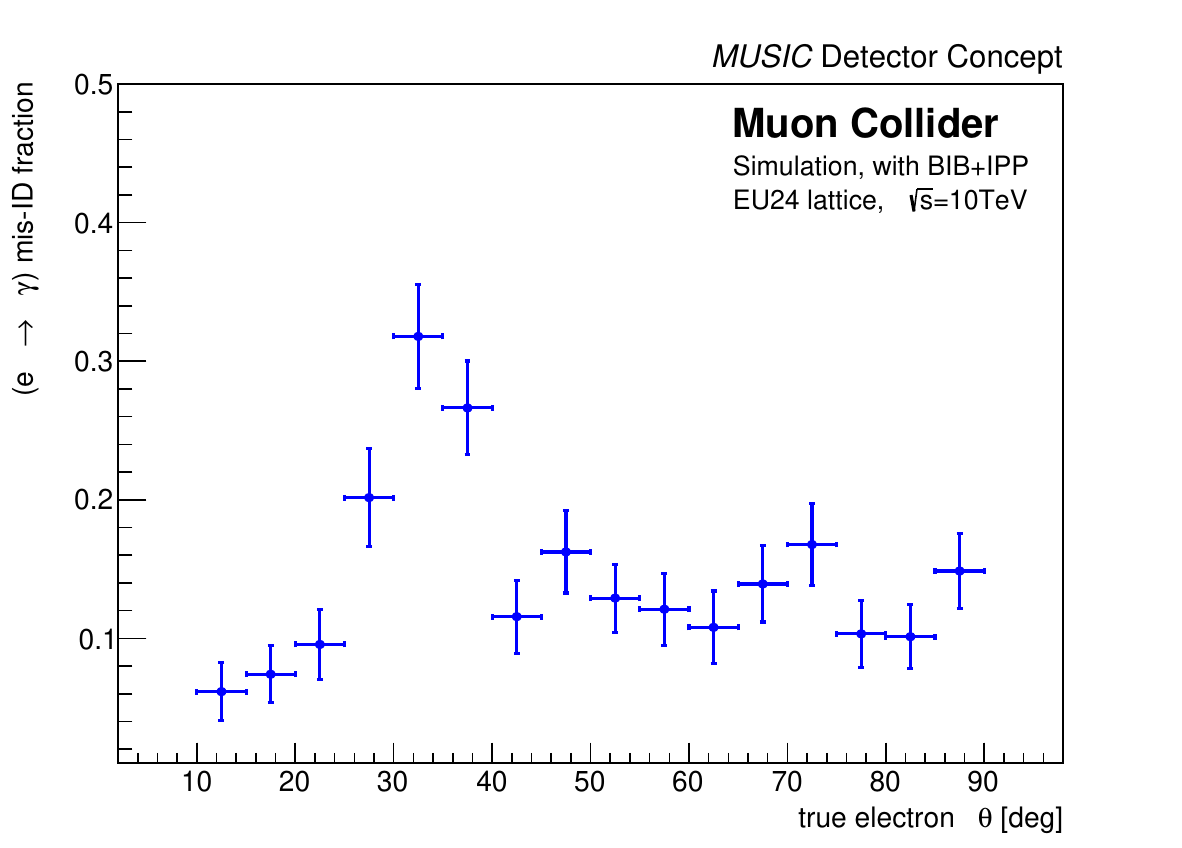}
    \caption{Electrons mis-identification probability to photons, as a function of Monte Carlo electron polar angle.}
    \label{fig:electron_misid}
\end{figure}

The mis-identification of photons to electrons is evaluated event-by-event, by matching ECAL clusters to the montecarlo-level photon, and checking whether the cluster is associated to an electron candidate. The probability of a photon being identified as an electron is $\sim 6\%$ on average, being peaked at small polar angles, and decreases with increasing photon energies (Figure \ref{fig:photon_misid}).

In an analogous way, the mis-identification of electrons to photons is evaluated event-by-event, by matching ECAL clusters to the montecarlo-level electron, and checking whether the cluster is associated to a photon candidate. The probability of an electron being identified as a photon is $\sim 14\%$ on average, slightly increasing for increasing electron's $p_T$, and is peaked at $\theta \sim 35^\circ$ (Figure \ref{fig:electron_misid}). This peculiar distribution stems mostly from the lower tracking efficiency in this region, that corresponds to the transition between the barrel and endcap sections of the tracker.

\begin{figure}
    \centering
     \includegraphics[width=1\linewidth]{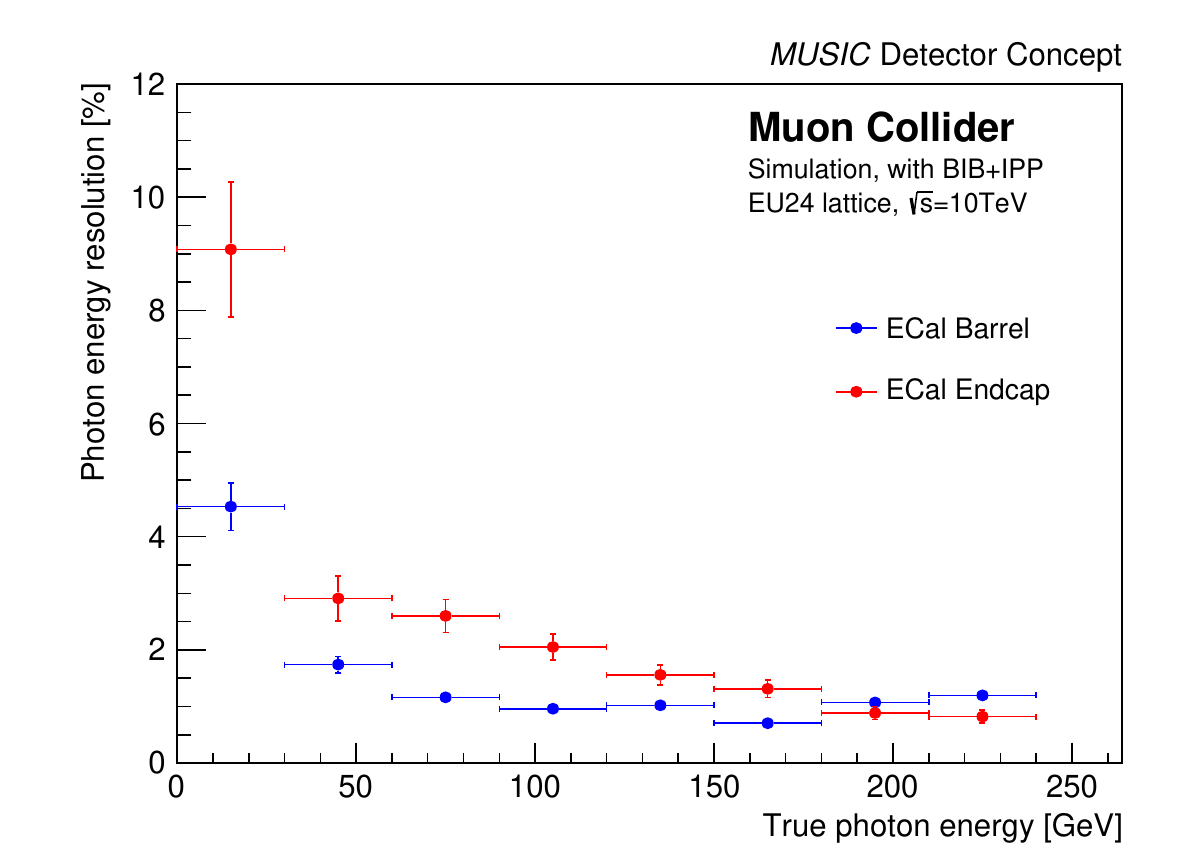}
    \caption{Photon energy resolution in ECAL barrel (blue points) and endcaps (red points) as a function of Monte Carlo photon energy.}
    \label{fig:photon_res}
\end{figure}

The photon energy resolution is evaluated in bins of true photon energy, $E_{MC}$. For each bin, the fractional residual $r=(E_{reco}-E_{MC})/E_{MC}$ is formed, and the RMS of the $r$ distribution is taken as the resolution estimate. The results in Fig.~\ref{fig:photon_res} indicate  $\Delta E/E \simeq 10\%/\sqrt{E \text{ [GeV]}}$ in the barrel and $\Delta E/E \simeq 17\%/\sqrt{E \text{ [GeV]}}$.
The endcap performance is degraded by a higher flux of background photons; mitigation is expected from an optimized nozzle design (in terms of geometry and materials) and from more advanced reconstruction algorithms, such as tighter timing, improved topological clustering, and refined shower-shape selections.

% An electron candidate is defined as a reconstructed track matched to an ECAL cluster that satisfies the quality requirements specified for tracks and clusters. If multiple clusters are associated with the same track, the cluster with the largest energy is retained as the electron candidate.
% Truth matching proceeds by searching for each generator-level electron, reconstructed candidates within a small angular cone
% $\Delta R=\sqrt{\Delta \theta^2+\Delta \phi^2} < 5\cdot 10^{-3}$. 
% If several candidates satisfy this condition, the one with the smallest $\Delta R$ is selected. A generator-level electron is counted as reconstructed if at least one candidate passes the matching criterion.

% The electron identification efficiency results from the convolution of tracking and ECAL cluster finding. Figure~\ref{fig:ele_eff} shows that it is around 90\% for $p_T>10$ GeV in the barrel, dropping to about 20\% at low momenta. 

\begin{figure}[ht]
    \centering
    \includegraphics[width=1\linewidth]{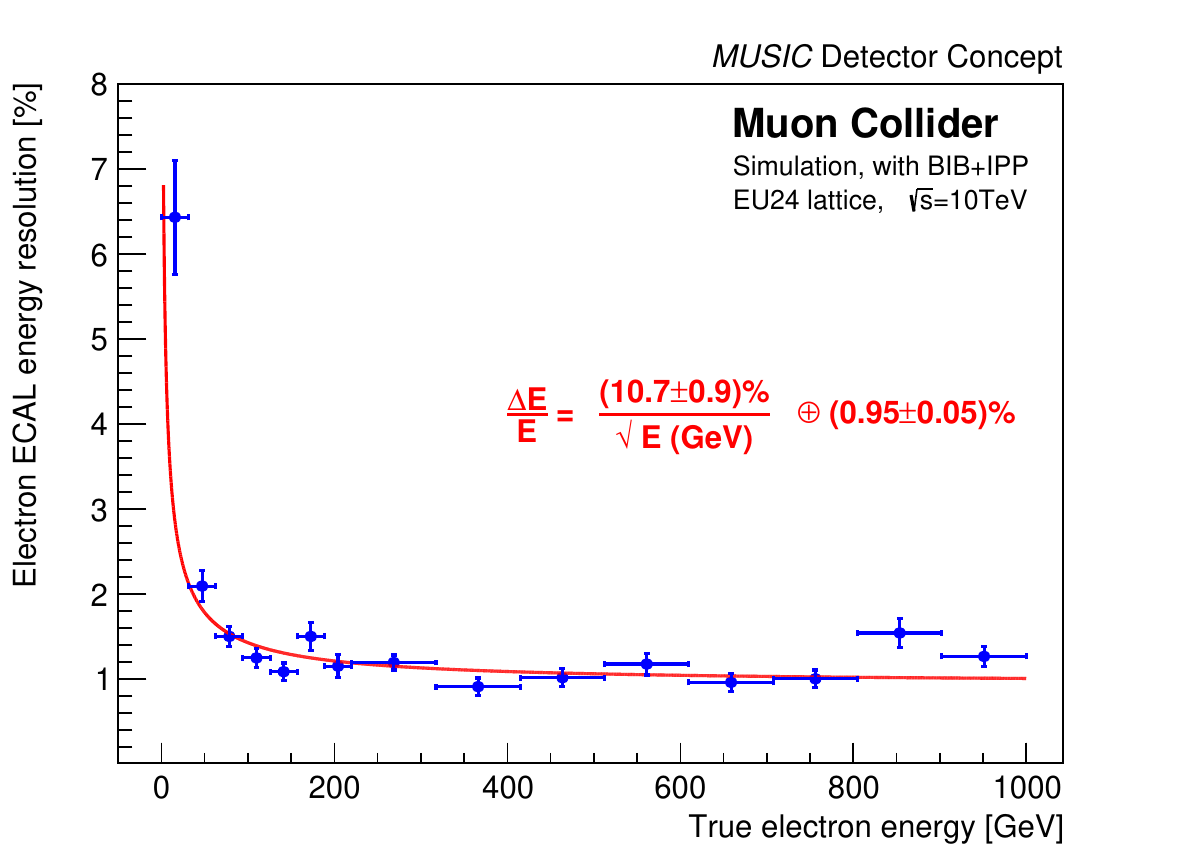}
    \caption{Electron energy resolution in ECAL barrel as a function of Monte Carlo electron energy.}
    \label{fig:ele_res}
\end{figure}

The energy resolution for electrons reconstructed in the ECAL barrel, shown in Fig.~\ref{fig:ele_res}, is consistent with that found for photons. In the ECAL endcaps, further developments in the track-cluster associations are needed to obtain an energy resolution which is compatible to photons, due to the larger contributions of backgrounds both in ECAL and at the track level.
At present, the electron reconstruction algorithm does not account for bremsstrahlung photon recovery, which impacts the energy resolution at low energies. Further improvements are expected upon the implementation of this feature.
The resolution distribution is fitted with the function $ \frac{\Delta E}{E} = \frac{a}{\sqrt{E}} \oplus b $ obtaining:
$$
    a = (10.7\pm0.9)\%\, , \qquad b = (0.95\pm0.05)\% \, .
$$

% The large track density from beam-induced background can lead to spurious associations between ECAL clusters and nearby tracks. Because photon showers closely resemble those of electrons, such mismatches typically lead to photon-to-electron misidentifications.
% The misidentification rate is estimated by reconstructing a sample of photons ($N_{tot}$) with the same configuration used for electrons and then performing truth matching. A reconstructed candidate is considered misidentified if, within $\Delta R < R_0$ (with the same $R_0$ as for the photon efficiency computation), no neutral ECAL cluster is reconstructed, but at least one ECAL cluster associated with a track is found. 
% Particles are binned into ten intervals of generator-level energy $E_{MC}$, and in each bin the misidentification fraction is $f_{mis}=N_{mis}/N_{tot}$. The dependence of $f_{mis}$ on $E_{MC}$ is shown in Fig.~\ref{fig:miside}.
% A future study will assess the development of a more optimized track-cluster association algorithm, exploiting more detailed track-cluster compatibility parameters to minimize the misidentification probability.

% %
% \begin{figure}[ht]
%     \centering
%     \includegraphics[width=1\linewidth]{figures/photMisId.pdf}
%     \caption{Fraction of photons misidentified as charged particles, as a function of the photon's $E_{MC}$. }
%     \label{fig:miside}
% \end{figure} 
%

\subsection{Jet reconstruction algorithm}
\label{sec:jets}
Jet reconstruction at a high-energy muon collider faces the challenge of unique machine-induced backgrounds, and therefore requires algorithms tailored to this environment. Developing a dedicated jet finder and background-mitigation strategy lies beyond the scope of this study and will require a focused effort. For the present analysis, jets are reconstructed with a tuned Particle-Flow configuration provided by PandoraPFA.

Tracks, ECAL clusters, and HCAL clusters are used as inputs to the Particle Flow algorithm.
Tracks are reconstructed as described in Sec.~\ref{sec:tracking} and further spurious-tracks removal procedures are specifically applied at the particle identification level to suppress the contamination from beam-induced background combinatorial tracks. Clusters in ECAL are selected following the prescriptions used for photons in Sec.~\ref{sec:calo}. HCAL cluster reconstruction proceeds with the clustering algorithm available in PandoraPFA, by requiring a minimum energy in the cell of 2 MeV in order to reduce the background.

The tracking and calorimetric information are combined by the Particle Flow algorithm to produce a list of charged and neutral particle candidates. Tracks are labelled as charged particles if they pass a set of requirements, while calorimeter clusters not associated with a track and passing other requirements are considered neutral particles. 
Jets are then clustered from the Particle Flow candidates using the $k_t$ algorithm~\cite{kt} with a radius parameter $R=0.5$, The $R$ parameter has been chosen among two different tested values (0.5 and 0.7) to obtain the best dijet mass resolution and efficiency for simulated and reconstructed Higgs boson decays to $b \bar{b}$ events.

The jet four-momentum is obtained by summing the four-momenta of its constituent particles. The jet axis is defined as the direction of the jet momentum.
Reconstructed jets are compared to truth-level jets, \emph{i.e.} jets clustered applying the $k_t$ algorithm with $R=0.5$ to the list of truth-level stable particles (excluding neutrinos) in the simulation, in order to evaluate corrections and to determine the jet performance. In particular, a jet is matched with a truth-level jet if the $\Delta R$ distance between their two axes is below 0.5.
A jet $p_T$ correction, obtained from a comparison between reconstructed jets and matched truth-level jets, is applied. 
%Additionally, a correction to the jet axis direction is implemented for jets in the transition regions between the barrel and endcap detectors, defined by $30^\circ< \theta < 60^\circ$ and $120^\circ< \theta < 150^\circ$, where the direction can be distorted by the high yield of combinatorial tracks from the beam-induced background.

The jet performance is evaluated using samples of events of $b\bar{b}$, $c\bar{c}$, and light-dijet events, generated with PYTHIA requiring an approximate flat distribution of the dijet invariant mass, simulated and reconstructed through the full pipeline.

Given the high number of tracks and clusters reconstructed from the machine-induced background combinatorial, a significant number of fake jets are expected, where a fake jet is defined as a jet not matched with a truth-level jet.
The contribution of the fake jets is minimized by requiring at least one track per jet, two neutral clusters per jet, and the fractional contribution of neutral clusters to the total $p_T$ greater than 5\%. In this way, an average number of 0.59 fake jets per bunch crossing is found. No additional cleaning is applied at this level,  as this number can be further reduced through jet flavor identification algorithms and/or analysis requirements. Notably, most of the fake jets are concentrated in the transition region between barrel and endcap detectors.

\begin{figure}[ht]
    \centering
    \includegraphics[width=1\linewidth]{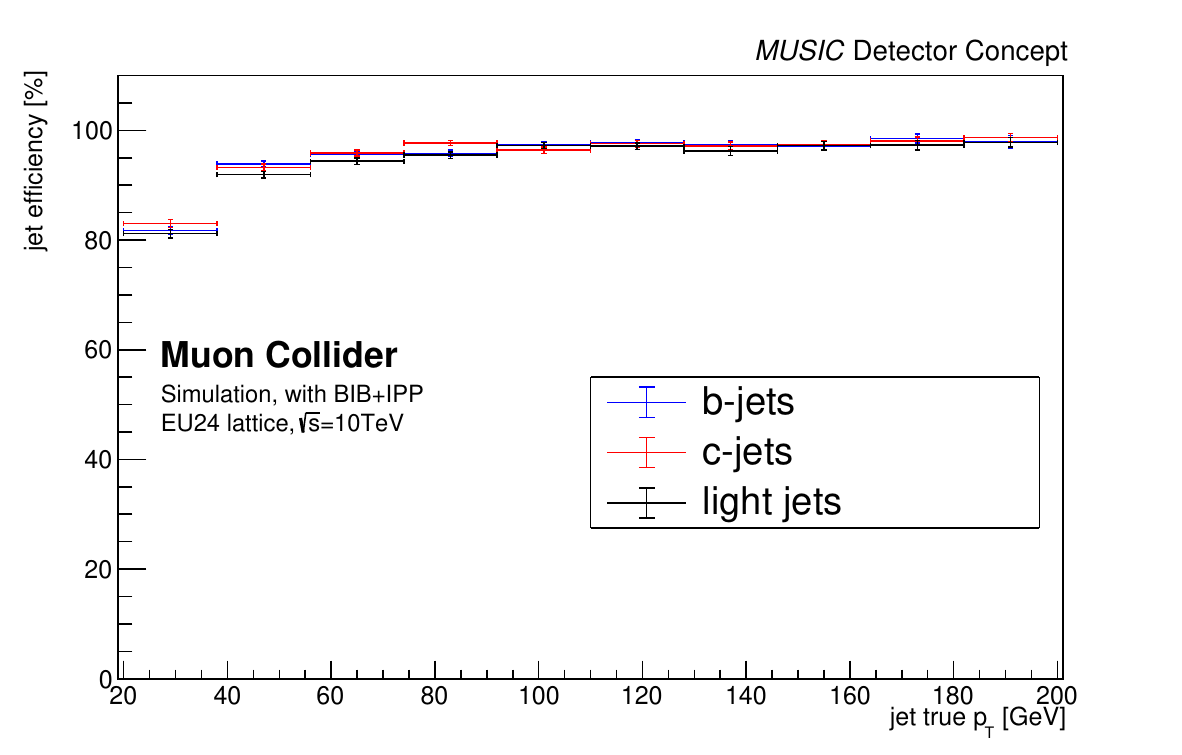}
    \caption{Jet reconstruction efficiency as a function of the Monte Carlo jet $p_T$ for different flavours of the quark originating the jet. }
    \label{fig:jet_eff}
\end{figure}

\begin{figure}[ht]
    \centering
    \includegraphics[width=1\linewidth]{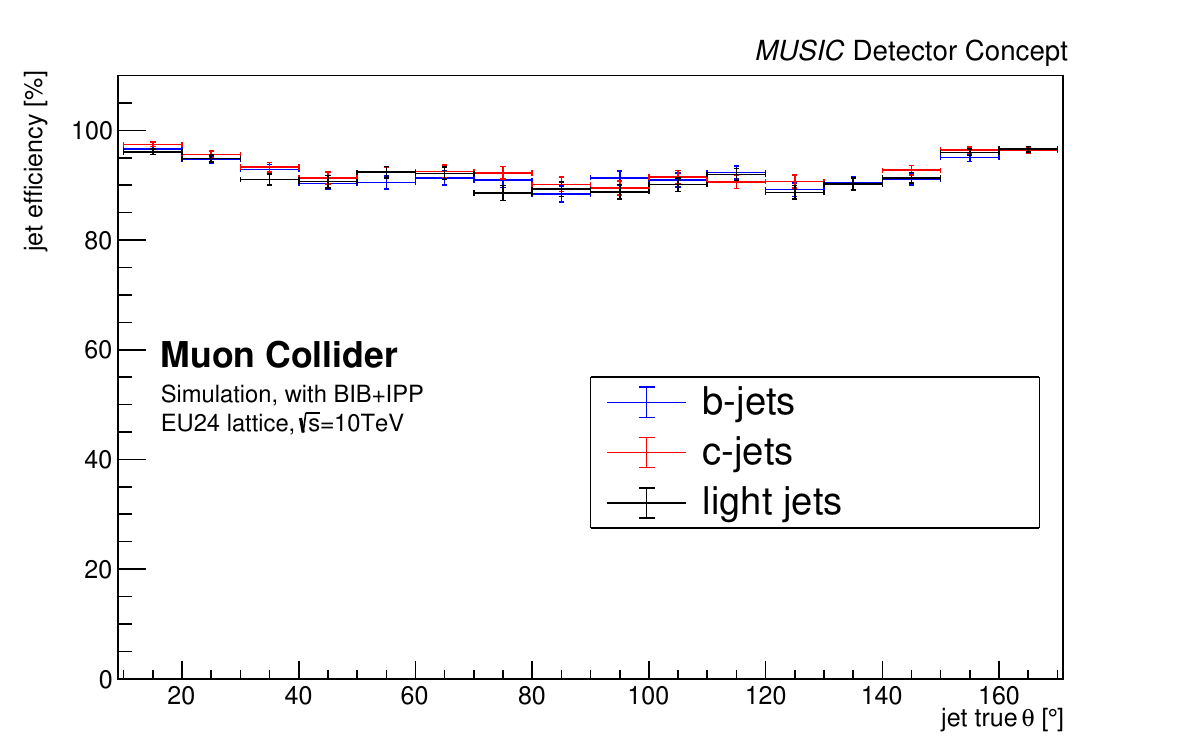}
    \caption{Jet reconstruction efficiency as a function of the Monte Carlo jet axis $\theta$ for different flavours of the quark originating the jet. }
    \label{fig:jet_eff_theta}
\end{figure}

The jet reconstruction efficiency is defined as $\epsilon_{\mathrm{jet}}=N^{\mathrm{reco}}_{\mathrm{jet}}/N^{\mathrm{true}}_{\mathrm{jet}}$ where $N^{\mathrm{reco}}_{\mathrm{jet}}$ is the number of reconstructed jets matched with a truth-level jet and $N^{\mathrm{true}}_{\mathrm{jet}}$ is the number of truth-level jets.
The jet reconstruction efficiencies are shown in Fig.~\ref{fig:jet_eff} as a function of the Monte Carlo jet momentum for different jet flavours and in Fig.~\ref{fig:jet_eff_theta} as a function of the Monte Carlo jet axis $\theta$. The efficiency is approximately 80$\%$ for jets with $p_T$ around 20 GeV and exceeds $95\%$ for $p_T$ above 60 GeV. The efficiency remains consistent across different jet flavors.
%In the central region, the efficiency is close to 100$\%$ and remains consistent across different jet flavors.

\begin{figure}[ht!]
    \centering
    \includegraphics[width=1\linewidth]{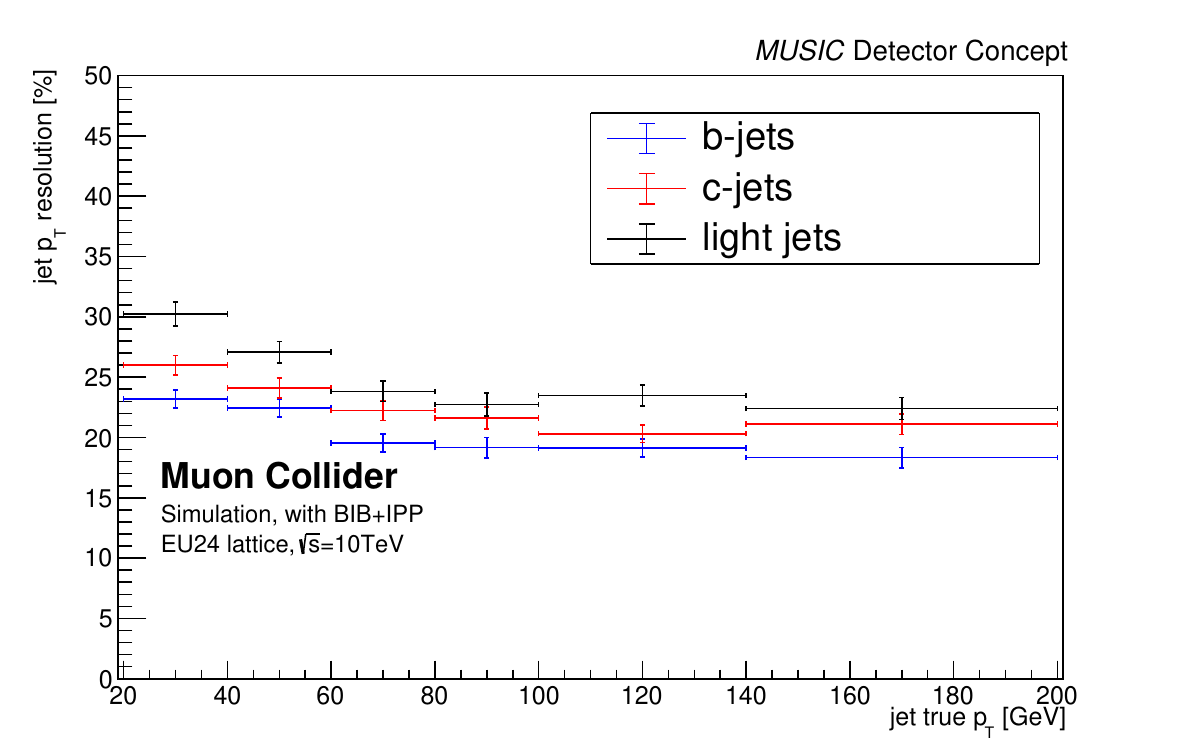}
    \caption{Jet transverse momentum resolution as a function of jet $p_T^{MC}$ for different flavours of the quark originating the jet in the central region $60^\circ<\theta<120^\circ$. }
    \label{fig:jet_res_cent}
\end{figure}

\begin{figure}[ht!]
     \includegraphics[width=1\linewidth]{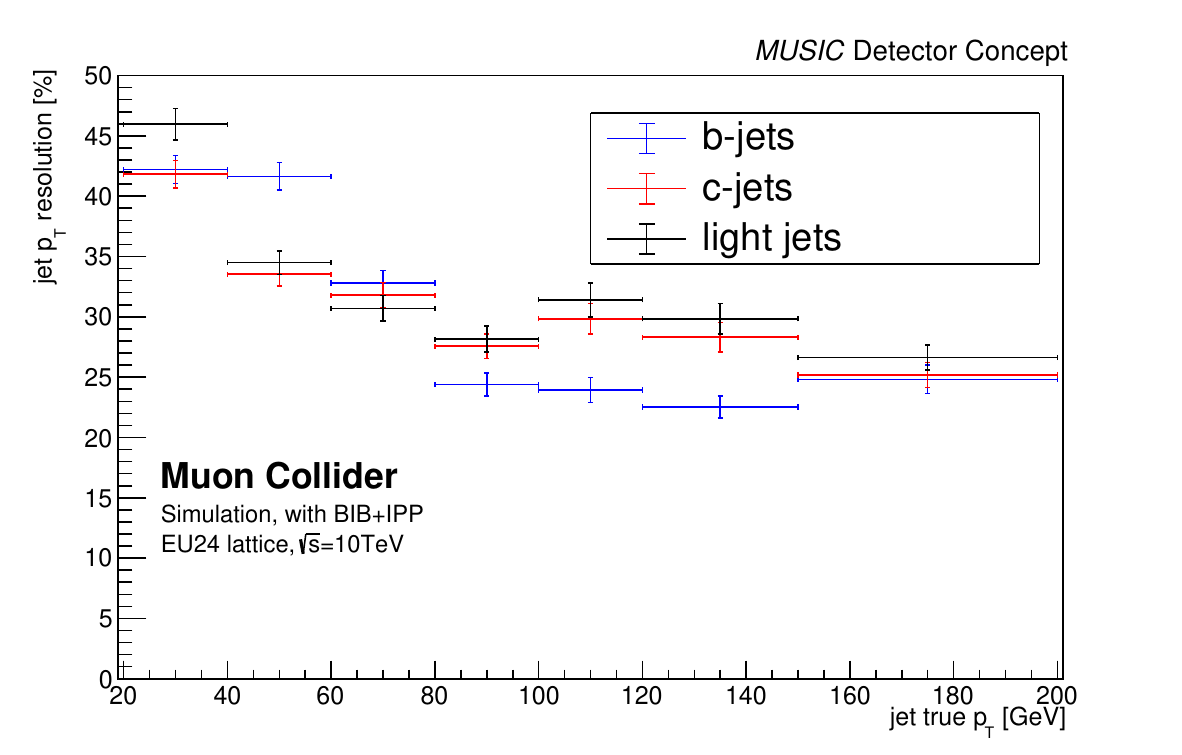}
    \caption{Jet transverse momentum resolution as a function of jet $p_T^{MC}$ for different flavours of the quark originating the jet in the barrel-endcap transition regions $30^\circ<\theta<60^\circ$ and $120^\circ<\theta<150^\circ$. }
    \label{fig:jet_res_trans}
\end{figure}

The jet $p_T$ resolution is defined as the RMS95 (root mean square of the smallest interval that contains 95\% of the events) of the $(p_T^{corr}-p_T^{MC})/p_T^{MC}$ distribution, where $p_T^{corr}$ is the corrected jet $p_T$ and $p_T^{MC}$ is the $p_T$ of the matched truth-level jet, and it is studied as a function of several variables. Figure~\ref{fig:jet_res_cent} shows the achieved resolution as a function of the $p_T^{MC}$ in the central region ($60^\circ<\theta<120^\circ$). For $p_T^{MC}\sim 20$ GeV, it is around 23$\%$ for $b$-jets and it improves to 19$\%$ at $p_T^{MC} \sim 200$ GeV.
In the transition regions, i.e. $30^\circ<\theta<60^\circ$ and $120^\circ<\theta<150^\circ$, the performance degrades at low $p_T^{MC}$ with the resolution ranging from 43$\%$ at $b$-jet $p_T^{MC} \sim 20$ GeV to 25$\%$ at $b$-jet $p_T^{MC} \sim 100$ GeV as documented in Fig.~\ref{fig:jet_res_trans}.

\begin{figure}[ht!]
     \includegraphics[width=1\linewidth]{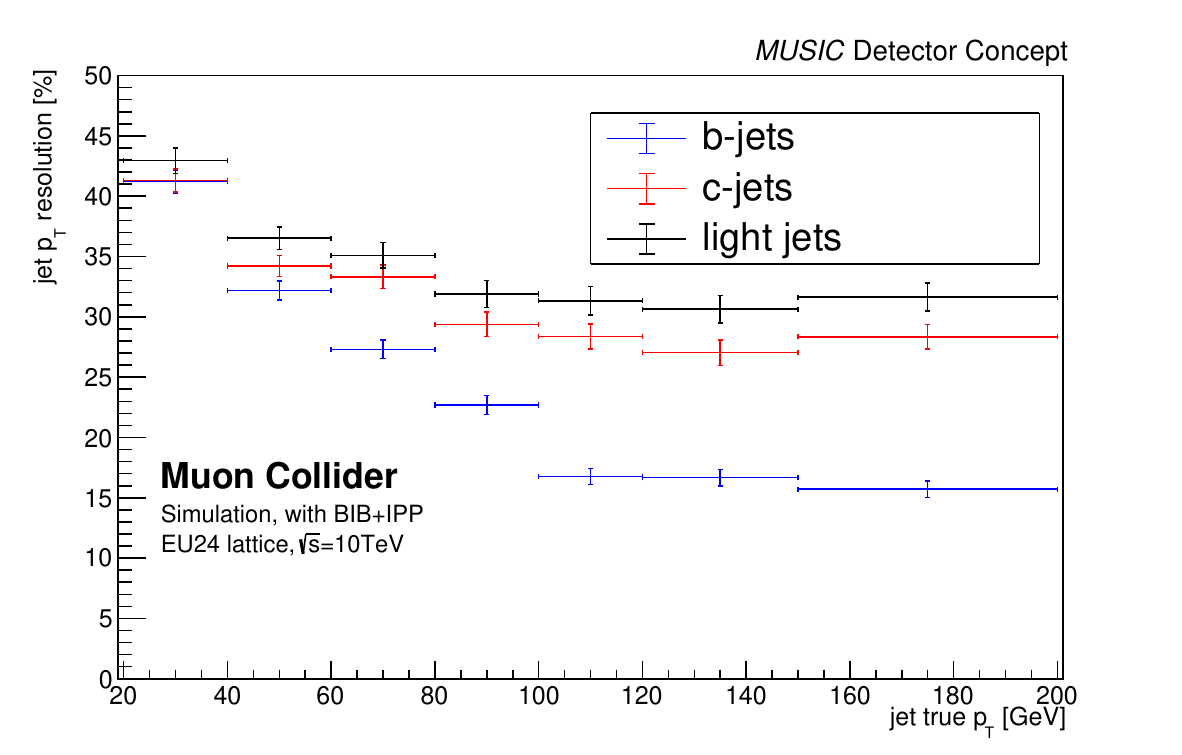}
    \caption{Jet transverse momentum resolution as a function of jet $p_T^{MC}$ for different flavours of the quark originating the jet in the forward regions $\theta<30^\circ$ and $\theta>150^\circ$. }
    \label{fig:jet_res_forw}
\end{figure}

In the forward regions ($\theta<30^\circ$ and $\theta>150^\circ$), the transverse momentum resolution varies from 41$\%$ at $b$-jet $p_T^{MC}\sim 20$ GeV to 15 $\%$ at $b$-jet $p_T^{MC}\sim 100$ GeV, as shown in Fig.~\ref{fig:jet_res_forw}.
Although the resolution is not at the level achieved at $e^+e^-$ colliders \cite{Boronat:2016tgd}, it has proven adequate for statistically distinguishing the $H \to b \bar{b}$ and $Z \to b \bar{b}$ components of the data sample, enabling the measurement of the Higgs properties at the target precision~\cite{3TEV}. Nevertheless, this represents only an initial optimization for the 10 TeV detector, and there is significant room for improvement, which will be explored in future studies.
 
\subsection{Muon identification}
\label{sec:muon}
\begin{figure}[ht]
 \centering
    \includegraphics[width=1\linewidth]{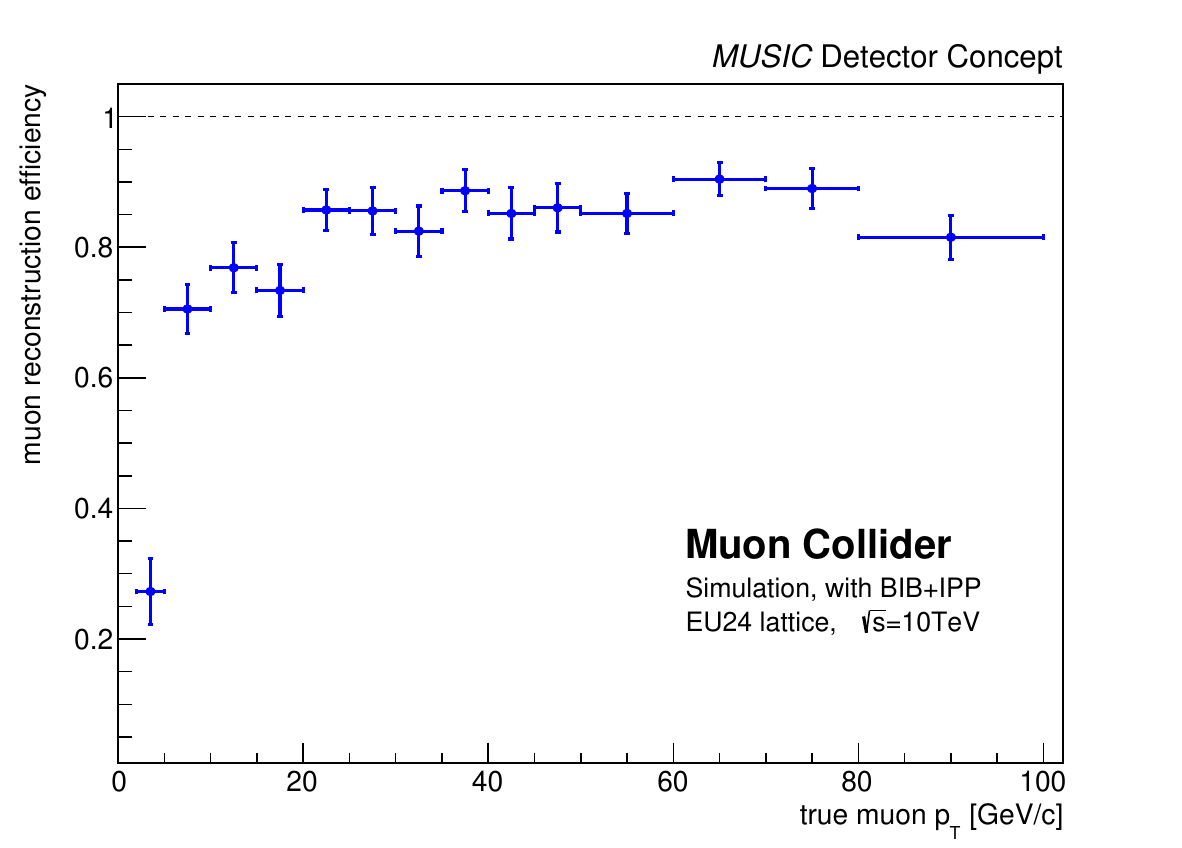} 
 \caption{Muon reconstruction and identification efficiency as a function of the muon's true $p_T$.}
 \label{fig:muons_effpt}
\end{figure}

\begin{figure}[ht]
 \centering
 \includegraphics[width=1\linewidth]{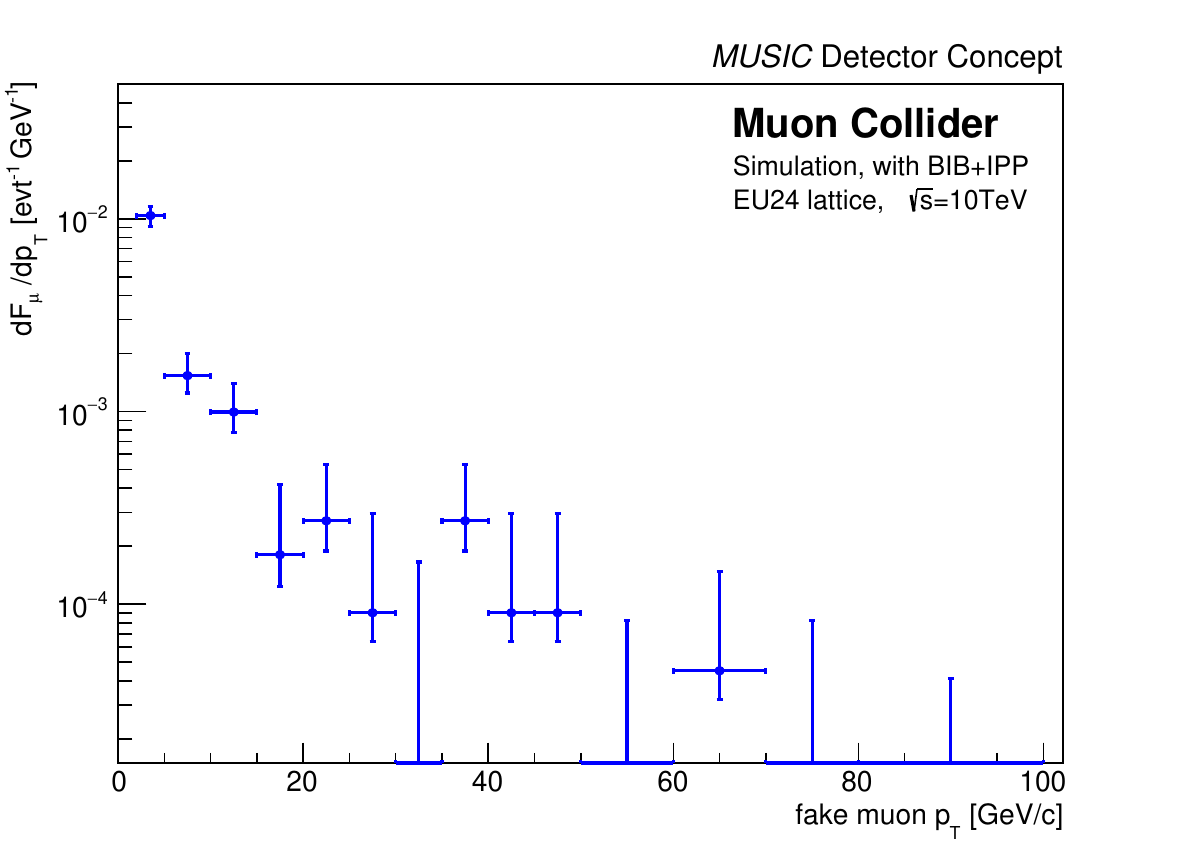}
 \caption{Differential muon fake rate, $F_\mu$, as a function of the fake muon's $p_T$. The error bars having no visible data point represent the rate upper limit for $p_T$ ranges where no fake muons are reconstructed in the studied sample.}
 \label{fig:muons_fakept}
\end{figure}

The muon reconstruction procedure is based on the association of a reconstructed track with hits in the calorimeters and the muon system. The digitization of the calorimeter \texttt{SimHits} is described in Sec.~\ref{sec:calo} (common for all reconstruction chains). The digitization of the muon system's \texttt{SimHits} proceeds by integrating the energies deposited by the particles for each detector cell in a single hit, and then applying selections on the recorded energy and time. While the energy threshold is the same for all cells, the time windows are optimized separately for three polar angle regions. The pattern recognition is performed via a dedicated PandoraPFA algorithm. Tracks are reconstructed as described in Sec.~\ref{sec:tracking} and further spurious-tracks removal procedures are specifically applied at the particle identification level to suppress the contamination from beam-induced background combinatorial tracks.
%optimized for the MUSIC detector, which takes as inputs the reconstructed tracks and the \texttt{DigiHits} of the Muon System, ECal, and HCal. 
First, the algorithm clusters the hits in the muon system, generating a set of candidate \textit{muon clusters}. This set undergoes a quality selection to remove most of the spurious clusters. The algorithm then extrapolates the tracks with $p_T > 2$ GeV in the magnetic field to the muon system. The next step is the matching between tracks and muon clusters. The matching parameters are the distance of closest approach between the track and the centroid of the cluster in the innermost layer of the muon system, and the reduced $\chi^2$ of the cluster's hits with respect to the track. If compatibility is found, a muon candidate is created. The algorithm then associates all calorimeter hits present along the track's extrapolated trajectory to the muon candidate. The reconstruction and identification parameters can be tuned to match the desired efficiency or muon fake rate, depending on the analysis context. Figures \ref{fig:muons_effpt} and \ref{fig:muons_fakept} show, respectively, the muon reconstruction efficiency and the muon fake rate for a representative configuration. The efficiency starts at $\sim 30\%$ for muons with $p_T < 5$ GeV, rising quickly to $\sim80\%$ and reaching a maximum of $\sim90\%$ for $p_T>40$ GeV. The muon fake rate shows a maximum for $p_T < 5$ GeV, decreases quickly for larger $p_T$, and becomes negligible for $p_T>15$ GeV. Integrating over the $p_T>2$ GeV range, fake muons arise with a rate of $\sim 0.05$ per event, reaching $\lesssim 0.01$ per event if requiring $p_T>10$ GeV.

\vspace{-20pt}

\subsection{Jet flavour tagging algorithm}
\label{sec:flavour}

\begin{figure}[ht]
 \centering
    \includegraphics[width=1\linewidth]{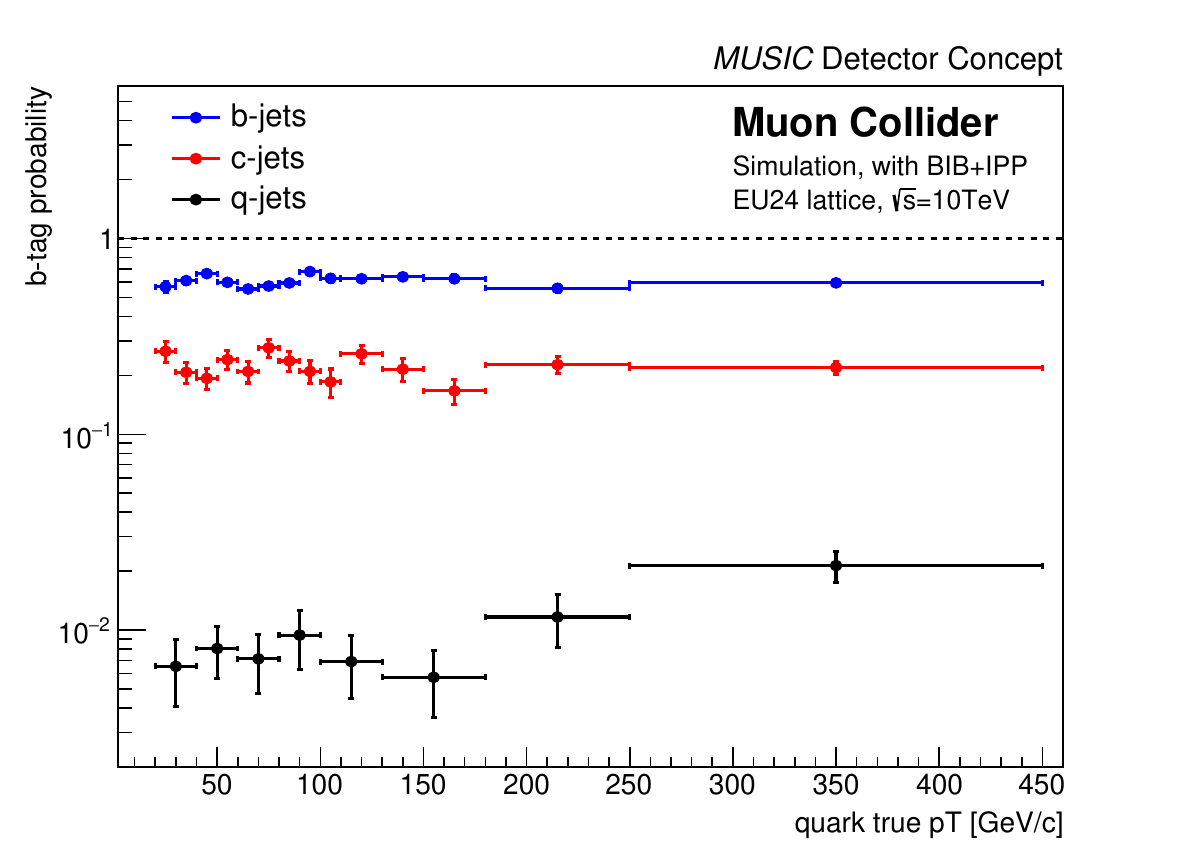} 
  
 \caption{Jet flavor identification efficiencies as a function of the quark true $p_T$ for \textit{b}, \textit{c}, and \textit{q} jets.}
 \label{fig:jet_tag_pt}
\end{figure}  

\begin{figure}[ht]
 \centering
 \includegraphics[width=1\linewidth]{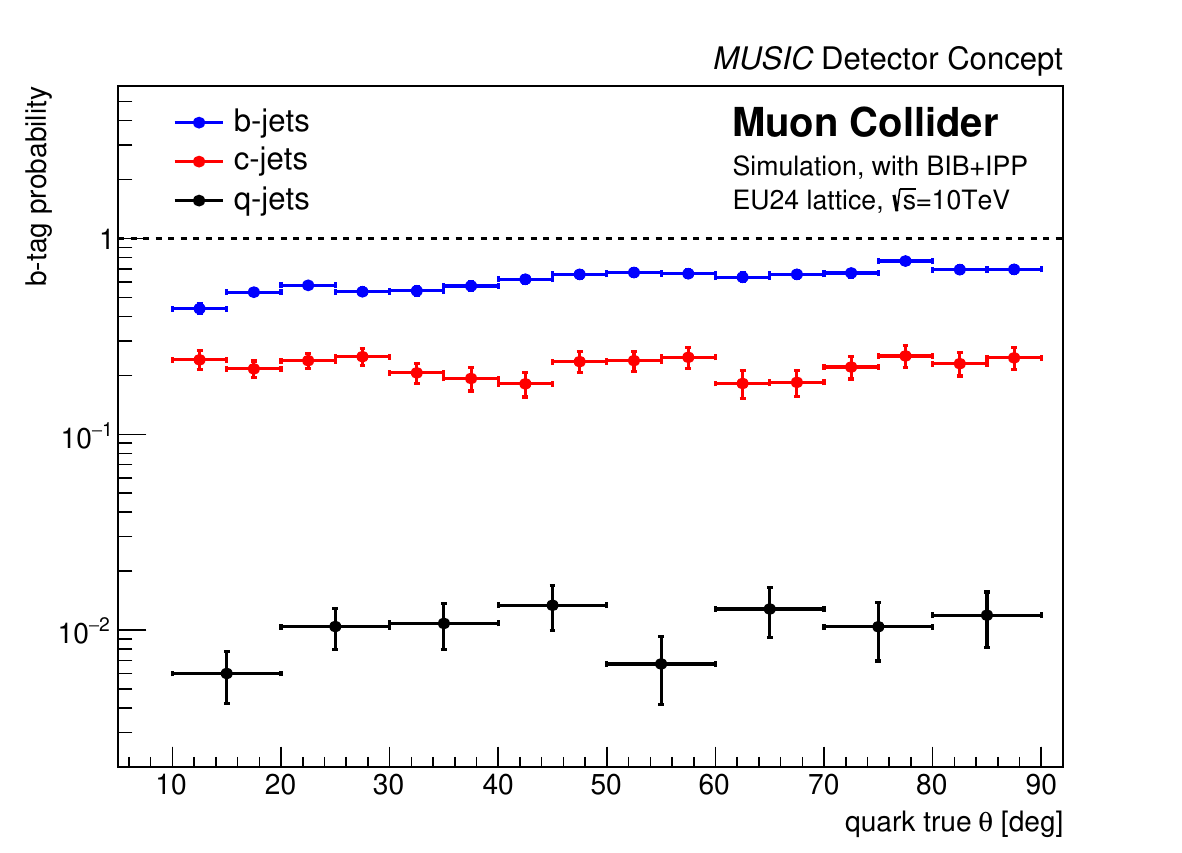}
 \caption{Jet flavor identification efficiencies as a function of the quark true $\theta$ for \textit{b}, \textit{c}, and \textit{q} jets.}
 \label{fig:jet_tag_theta}
\end{figure}

The jet flavour identification makes use of displaced vertices and in-jet muons as signatures of heavy-flavour hadron decays. A reconstructed secondary vertex signals the presence of long-lived $b$ or $c$ hadrons, while a muon within the jet cone strengthens the identification of semileptonic heavy-hadron decays. This study focuses on the identification of jets originating from $b$ quarks, while $c$-jet identification will be addressed in future studies. 

The algorithm follows the steps outlined below:
\begin{enumerate}
\item Jets are reconstructed according to the procedure described in Sec.~\ref{sec:jets}, and muons within the jet cone are identified following the criteria in Sec.~\ref{sec:muon}. Muons with transverse momentum $p_T > 2$ GeV are considered.
\item To improve the impact-parameter resolution, track reconstruction is repeated including the first layer of the vertex detector. Only hits within a distance $R = \sqrt{\Delta\eta^2+\Delta\phi^2}<1.0$ from the jet axis are considered (where $\eta$ indicates the pseudorapidity). The resulting set of tracks is then provided as input to a vertex-finding algorithm based on the LCFIplus package~\cite{lcfiplus}, specifically optimized for the muon-collider environment.
Candidate secondary vertices (SVs) are required to satisfy $L<50$ mm, where $L$ is the three-dimensional distance between the SV and primary vertex, and \mbox{$L/\sigma_L>2$}, ensuring a significant SV displacement from the primary vertex. 
\end{enumerate}
Jets are classified into four exclusive categories based on the presence of a muon and/or a reconstructed secondary vertex:
\begin{itemize}
\item[\bf C0:] No secondary vertex and no muon associated with the jet.
\item[\bf C1:] A muon satisfying the above selection criteria, but no SV within the jet.
\item[\bf C2:] At least one SV fulfilling the above requirements, but no associated muon. In this category, jets are further classified using two Deep Neural Networks (DNNs): one trained to discriminate \textit{b}-jets from light-flavour jets, and another to separate \textit{b}-jets from \textit{c}-jets. Both networks are trained on 30 input variables derived from jet and SV properties. The variables include, as a non-exhaustive list: transverse and 3D displacement, significance, proper time of flight, and invariant mass of the SV, number and displacement of the SV-associated tracks, reconstructed $p_T$ of the SV, the fraction of the jet's $p_T$ carried by the SV, and combinations of the above. Depending on the analysis needs, different DNN working points can be applied.
\item[\textbf{C3:}] Both a muon and at least one SV, each satisfying the corresponding criteria, are found within the jet.
\end{itemize}

Jets belonging to categories \textbf{C1}, \textbf{C2}, or \textbf{C3} are considered \textit{tagged}. No additional selection is imposed at this stage; further refinement and optimization are left to individual analyses, where event topology, kinematics, and physics goals can be fully taken into account.

The performance of the $b$-flavour tagging algorithm is assessed on simulated $b\bar{b}$, $c\bar{c}$, and light-quark ($q\bar{q}$) dijet samples. The events are generated such that the dijet invariant mass is approximately flatly distributed, allowing for a uniform evaluation of the tagging efficiency and misidentification probability over a broad kinematic range.

The $b$-jet identification efficiency, as well as the probabilities for $c$ and light-quark jets to be misidentified as $b$ jets, are shown in Figs.~\ref{fig:jet_tag_pt} and \ref{fig:jet_tag_theta} as functions of the true quark transverse momentum $p_T$ and polar angle $\theta$.
%The flavor identification efficiency for $b$ jets and the probabilities of $c$ and light-quark jets to be misidentified as $b$ jets are shown in Figs.~\ref{fig:jet_tag_pt} and \ref{fig:jet_tag_theta} as a function of the true quark transverse momentum, $p_T$, and polar angle, $\theta$. 
The reference working point yields an average efficiency of $60\%$ for \textit{b}-jets, and average misidentification probabilities of $22\%$ for \textit{c}-jets, and $0.9\%$ for \textit{q}-jets.
Under these conditions, the \textit{b}-tagging efficiency remains nearly constant between $60$ and $70\%$ in the central angular region, decreases smoothly to about $\sim\!55\%$ in the small-$\theta$ region, and reaches $\sim\!45\%$ near the acceptance boundaries. It remains constant around the average $60\%$ value through the $p_T$ range.

The \textit{c}-jet misidentification probability is nearly constant between $20\%$ and $25\%$ in the full ranges of $p_T$ and polar angle.
The light-jet misidentification probability remains low and uniform around $\sim\!1\%$ through the angular range, while showing a mild increase from $\sim\!0.6\%$ at low transverse momenta to $2\%$ for $p_T > 200$ GeV.

\section{Summary and outlook}

The proof-of-concept design of the MUSIC detector presented in this work introduces a tracking system optimized for robust performance in the forward region, where machine-induced backgrounds have the most significant impact on tracking efficiency and purity. Similarly, the CRILIN calorimeter is proposed to minimize the effects of this background contamination in the innermost layers of the electromagnetic calorimeter.
The origin of machine-induced backgrounds is examined in detail for the tracker and electromagnetic subsystems, which are the most affected components. This preliminary study will play a crucial role in developing the Conceptual Design Report of the detector and the machine interaction region, guiding the optimization of magnet positioning, shielding geometry, and detector layout to reduce the flux of background particles entering the detector. A detailed study is foreseen in future work, to be carried out in close coordination with accelerator experts for the design and optimization of the interaction region.
A preliminary discussion of the detector magnet and its corresponding magnetic field map is presented, although a uniform magnetic field is assumed in the current event reconstruction. In the subsequent phases of sub-detector design and event reconstruction studies, the detailed magnetic field map will be implemented to provide a more realistic evaluation of the detector performance.
The performance of physics object reconstruction and identification has been evaluated using the current detector and IR configuration, accounting for the effects of machine-induced backgrounds. The software algorithms have been tuned to cope with the high particle fluxes and to mitigate their impact; however, they are not yet fully optimized. Future developments envision a substantially broader integration of artificial intelligence–based algorithms. Both track and jet reconstruction are expected to significantly benefit from these approaches, as already demonstrated by the LHC experiments. Nevertheless, strong performance is achieved. Tracks, electrons, photons, and muons are reconstructed with efficiencies exceeding overall $90\%$. The spurious hits generated by machine-induced backgrounds cause a modest reduction in purity at small angles from the beam direction, while their impact is negligible in the central region. Jet reconstruction and identification have reached a high level of maturity, with efficiency above $80\%$ for jet $p_T$ greater than 20 GeV and $90\%$ for jet $p_T$ greater than 40 GeV, with a fake rate of less than one fake jet per bunch crossing. This result is particularly notable, since jet reconstruction is affected by the machine-induced background through both the tracker and calorimeter systems.
The region below $10^\circ$, currently occupied by the nozzle, is not instrumented and thus constitutes a dead zone for MUSIC. 
Future detector designs will include studies aimed at instrumenting this region. In addition to increasing the overall detector acceptance, such an improvement would enable the tagging of forward muons, providing sensitivity to processes occurring at small polar angles. The identification of these forward muons will be essential for distinguishing between $W$ boson fusion from $Z^0$ boson fusion mechanisms, thereby enhancing the capability to study electroweak symmetry breaking and vector-boson scattering processes. Moreover, an extended instrumentation in this angular region could play a significant role in determining the luminosity, compensating for the absence of dedicated sub-detectors currently included in the baseline design.

\section*{Acknowledgments}

We are grateful to the International Muon Collider Collaboration for their support. 
We acknowledge the financial support of the Italian National Institute for Nuclear Physics (INFN), the University of Padua, 
and the European Organization for Nuclear Research (CERN). CloudVeneto is acknowledged for the use of computing and storage facilities. 
This work was supported by the European Union’s Horizon 2020 and Horizon Europe Research and
Innovation programs through the Marie Sk\l{}odowska-Curie RISE Grant Agreement No. 101006726, the Research Infrastructures 
INFRADEV Grant Agreement No. 101094300, and the EXCELLENT SCIENCE - Research Infrastructures Research Innovation Grant Agreement No. 101004761 and Grant Agreement No. 101004730.

\bibliography{sn-bibliography}% common bib file
\end{document}